\documentclass[a4paper,twoside,openany,11pt]{memoir} 
\pdfoutput=1

\def\fullheadfoot{0}
\usepackage[english]{babel}
\usepackage{microtype,xspace}
\usepackage[utf8]{inputenc}	

\usepackage{amsfonts,amsmath,amssymb,bm,mathrsfs,mathtools,dsfont} 	
\allowdisplaybreaks 	
\usepackage{xcolor}
\usepackage{hyperref}
\usepackage{slashed}
\usepackage{multicol}
\usepackage{fontawesome}

\usepackage{siunitx}
\usepackage{listings}
\definecolor{bg}{RGB}{247, 247, 247}
\definecolor{frame}{RGB}{207, 207, 207}
\definecolor{stringgray}{RGB}{100, 100, 100}
\definecolor{vargreen}{RGB}{60, 135, 40}
\definecolor{comment}{RGB}{50, 150, 200}

\lstnewenvironment{function}[1][]{\needspace{3\baselineskip} \lstset{numbers = none, 
		frame = {lines}, rulecolor = \color{black}, xleftmargin=0pt, xrightmargin=0pt, framexleftmargin=0mm, framexrightmargin=0mm, backgroundcolor = \color{white}, #1}}{\vspace{-.5em}}
\newcommand{\argument}[1]{\lstinline[emph={#1}]|#1|}

\newenvironment{arguments}{\vspace{.2em} \begin{itemize}[label={}, itemsep = -.3em, itemindent = -1.5em] \vspace{-.8em}}{\end{itemize}}
\newcommand{\options}{\item \lstinline|Options| \vspace{-.5em}}
\newcommand{\argend}{\\[-.8em] 
}

\lstnewenvironment{example}[1][]{\lstset{numbers = none, framexleftmargin=0mm, xleftmargin=0pt, xrightmargin=0pt, backgroundcolor = \color{white}, frame = none, #1}}{}

\usepackage{geometry}
\usepackage[sort&compress,numbers,merge]{natbib}
\usepackage{chapterbib}
\makeatletter
\renewcommand{\@memb@bchap}{ 
\bibmark \prebibhook
}
\makeatother

\usepackage{graphicx}
\graphicspath{{./Figures/}}
\usepackage{caption,subcaption}
\usepackage{multirow}
\renewcommand{\arraystretch}{1.2}
\usepackage{tabularx}
\newcolumntype{Y}{>{\centering\arraybackslash}X}

\usepackage[shortlabels]{enumitem}
\setlist{itemsep=-.3em,topsep=.5em}
\SetEnumerateShortLabel{i}{\textit{\roman*}}

\definecolor{red}{rgb}{0.6,.0706,.1373}
\definecolor{blue}{rgb}{0,0.396,0.741}
\colorlet{blueRef}{blue!80!black}
\colorlet{blueLink}{blue!100!black}
\hypersetup{
	colorlinks, 
	bookmarksopen, 
	bookmarksnumbered,
	citecolor=blueLink, 	
	linkcolor=blueLink,	
	urlcolor=blueLink,			
	pdftitle={RGBeta},    
	pdfsubject={Renormalization group functions},	
	pdfauthor={Anders Eller Thomsen},    	
}
\PassOptionsToPackage{hyphens}{url}\usepackage{hyperref}

\addto\captionsenglish{}

\renewcommand{\contentsname}{Contents}
\renewcommand{\printtoctitle}[1]{}

\settocdepth{subsection}
\newcommand{\app}[1][Appendices]{
	\renewcommand{\thesubsection}{\Alph{subsection}}
	\numberwithin{equation}{subsection}
	\numberwithin{figure}{subsection}
	\pagestyle{appstyle}
	\sectionlike{#1} 
}

\makeatletter
\newcommand*\ifthispageodd{%
  \checkoddpage
  \ifoddpage
    \expandafter\@firstoftwo
  \else
    \expandafter\@secondoftwo
  \fi
}
\makeatother

\usepackage[noindentafter]{titlesec} 
\counterwithout{section}{chapter} 
\counterwithout{figure}{chapter} 
\counterwithout{table}{chapter} 
\numberwithin{equation}{section} 
\setsecnumdepth{subsubsection}

\DeclareMathAlphabet{\mathsfit}{OT1}{lmss}{m}{sl}
\DeclareMathAlphabet{\mathsfbf}{OT1}{lmss}{bx}{n}
\DeclareMathAlphabet{\mathsfbfit}{OT1}{lmss}{bx}{sl}
 
\DeclareMathVersion{chaptermath}
\SetSymbolFont{operators}{chaptermath}{OT1}{cmbr}{m}{n}
\SetSymbolFont{letters}{chaptermath}{OML}{cmbrm}{b}{it}
\SetSymbolFont{symbols}{chaptermath}{OMS}{cmbrs}{m}{n}
\DeclareMathVersion{subsectionmath}
\SetSymbolFont{operators}{subsectionmath}{OT1}{cmbr}{m}{n}
\SetSymbolFont{letters}{subsectionmath}{OML}{cmbrm}{m}{it}
\SetSymbolFont{symbols}{subsectionmath}{OMS}{cmbrs}{m}{n}

\titleformat{\section}{\centering \Large \bfseries \sffamily \mathversion{chaptermath} \color{blue!90!black} }{\thesection}{15pt}{}{}
\titlespacing{\section}{0pt}{15pt}{5pt}
\titleformat{\subsection}{\large \sffamily \mathversion{subsectionmath} \color{blue!90!black} }{\thesubsection}{10pt}{}{}
\titlespacing{\subsection}{0pt}{10pt}{5pt}
\titleformat{\subsubsection}{\normalsize \sffamily \itshape \mathversion{subsectionmath} \color{blue!80!black} }{\thesubsubsection}{10pt}{}{}
\titlespacing{\subsubsection}{0pt}{10pt}{0pt}

\newcommand{\sectionlike}[1]{\phantomsection \addcontentsline{toc}{section}{#1} \sectionmark{#1}
		\begin{center}
		\needspace{8\baselineskip}
		\Large \bfseries \sffamily \mathversion{chaptermath} \color{blue!90!black} #1  
		\end{center}
	\vspace{-5pt} 
}

\ifnum\fullheadfoot=1
	\makepagestyle{standardstyle}
	\makeoddhead{standardstyle}{\sffamily \mathversion{subsectionmath} \rightmark}{}{\sffamily \bfseries{\thepage}}
	\makeevenhead{standardstyle}{\sffamily \bfseries{\thepage}}{}{\sffamily \mathversion{subsectionmath} \leftmark}
	\makeheadrule{standardstyle}{\textwidth}{\normalrulethickness}
	\makepsmarks{standardstyle}{
		\nouppercaseheads
		\createmark{section}{both}{shownumber}{\ }{\hspace{1.2em}  }
		\createmark{subsection}{right}{shownumber}{}{\hspace{1.2em} }
	}
	
	\copypagestyle{appstyle}{standardstyle}
	\makeevenhead{appstyle}{\sffamily \bfseries{\thepage}}{}{ \sffamily \mathversion{subsectionmath} \leftmark}
	\makepsmarks{appstyle}{
		\nouppercaseheads
		\createmark{section}{both}{nonumber}{\ }{\ \ }
		\createmark{subsection}{right}{shownumber}{}{\ \ }
	}
\else 
	\makepagestyle{standardstyle}
	\makeoddfoot{standardstyle}{}{\sffamily -- {\bfseries\thepage} --}{} 
	\makeevenfoot{standardstyle}{}{\sffamily -- {\bfseries\thepage} --}{}
	\copypagestyle{appstyle}{standardstyle}
\fi

\createplainmark{toc}{both}{\contentsname}
\createplainmark{bib}{both}{\bibname}

\makeatletter
\let\MyIntOrig\int
\def\MyIntSpace{\hspace{-.35em}} 
\def\int{\MyInt}
\def\MyInt{\MyIntOrig\MyIntSkipMaybe}
\def\MyIntSkipMaybe{
	\@ifnextchar_{\MyIntSkipScript}{%
		\@ifnextchar^{\MyIntSkipScript}{%
			\@ifnextchar\limits{\MyIntSkipTok}{%
				\@ifnextchar\nolimits{\MyIntSkipTok}{%
					\MyIntSpace}}}}%
}
\def\MyIntSkipScript#1#2{#1{#2}\MyIntSkipMaybe}
\def\MyIntSkipTok#1{#1\MyIntSkipMaybe}

\newcommand{\pushright}[1]{\ifmeasuring@#1\else\omit\hfill$\displaystyle#1$\fi\ignorespaces}
\makeatother

\newcommand{\braces}[1]{\left\lbrace #1 \right\rbrace}

\newcommand{\eminus}{\vcenter{\hbox{\scalebox{0.6}[1]{$ - $}}}}	
\newcommand{\ord}[1]{\mathcal{O}\!\left( #1 \right)}

\newcommand{\hc}{\; + \; \mathrm{H.c.} \;}
\newcommand{\andeq}{\quad \mathrm{and} \quad}

\newcommand{\dd}{\mathop{}\!\mathrm{d}}
\newcommand{\ud}[2]{\phantom{}^{#1}\phantom{}_{#2}}
\newcommand{\du}[2]{\phantom{}_{#1}\phantom{}^{#2}}

\newcommand{\rep}[1]{\mathbf{#1}}
\newcommand{\repbar}[1]{\overline{\mathbf{#1}}}

\makeatletter
\newcommand{\vast}{\bBigg@{3}}
\makeatother

\renewcommand{\L}{\mathcal{L}}
\newcommand{\LL}{\mathrm{L}}
\newcommand{\RR}{\mathrm{R}}
\newcommand{\U}{\mathrm{U}}
\newcommand{\SU}{\mathrm{SU}}
\newcommand{\Sp}{\mathrm{Sp}}
\newcommand{\SO}{\mathrm{SO}}

\newcommand{\bef}{$ \beta $-function\xspace}
\newcommand{\befs}{$ \beta $-functions\xspace}
\newcommand{\msbar}{$ \overline{\text{\small MS}} $\xspace}
\newcommand{\rgbeta}{\texttt{RGBeta}\xspace}
\renewcommand{\quote}[1]{``#1''}
\newcommand{\hsc}[1]{{\normalsize\MakeUppercase{#1}}}

\begin{document}

\thispagestyle{empty}
\renewcommand*{\thefootnote}{\fnsymbol{footnote}}
\vspace*{0.05\textheight}
	{\sffamily {\bfseries \Huge \mathversion{chaptermath} \noindent RGBeta} \hfill }\\[-.7em]
	\textcolor{blue!80!black}{\rule{\textwidth}{3pt}}\\[.1em]
	{\sffamily \LARGE \mathversion{subsectionmath} A Mathematica Package for the Evaluation of Renormalization Group $ \beta $-Functions}
	
\begin{center}	
	\vspace{.015\textheight}
	{\sffamily \Large A\hsc{nders} E\hsc{ller} T\hsc{homsen}\footnote{thomsen@itp.unibe.ch}}
	\\[.2em]{ \sffamily \mathversion{subsectionmath} 
	Albert Einstein Center for Fundamental Physics, Institute for Theoretical Physics,\\[-.3em]
	University of Bern, CH-3012 Bern, Switzerland
	}
	\\[.005\textheight]{\itshape \sffamily \today}
	\\[.03\textheight]
\end{center}
\setcounter{footnote}{0}
\renewcommand*{\thefootnote}{\arabic{footnote}}%
\suppressfloats	

\begin{abstract}\vspace{-.03\textheight}
In completely generic four-dimensional gauge-Yukawa theories, the renormalization group $ \beta $-functions are known to the 3--2--2 loop order in gauge, Yukawa, and quartic couplings, respectively. It does, however, remain difficult to apply these results to specific models without the use of dedicated computer tools. We describe a procedure for extracting $ \beta $-functions using the general results and introduce \rgbeta, a dedicated Mathematica package for extracting the $ \overline{\text{\small MS}} $ $ \beta $-functions in broad classes of models. The package and example notebooks are available from the \texttt{GitHub} repository \href{https://github.com/aethomsen/RGBeta}{\faicon{github}}.
\end{abstract}
\vspace{.015\textheight}


\section{Introduction}
The renormalization group (RG) functions are fundamental quantities in quantum field theories (QFTs), governing how the dynamics of models change with energy. 
The \befs, in particular, determine the flow of couplings $ g $ with the renormalization scale $ \mu $. 
They are a staple of BSM physics used in grand unification to extrapolate low-energy couplings to the unification scale or to generate mass spectra from high-scale input in supersymmetric models. 
Other applications involve the study of the ultimate ultraviolet fates of models in the search of fundamental theories (see e.g. Refs.~\cite{Giudice:2014tma,Litim:2014uca}). 
It is, therefore, not surprising that the \befs were computed to the first few orders in perturbation theory several decades ago. 
In the general case of four-dimensional gauge-Yukawa theories, they were calculated to the 2-loop order in both gauge, Yukawa, and quartic couplings already in the 80's~\cite{Machacek:1983tz,*Machacek:1983fi,*Machacek:1984zw,Jack:1984vj}. What is perhaps more surprising is that it took another 20 years for the 3-loop result for the gauge \bef~\cite{Pickering:2001aq}, which represents the current state-of-the-art in generic theories. 

One contributing cause for this measured pace might be that although the general results for the \bef have long been known, it is an often time-consuming endeavor to apply the results to specific models. Typically, model builders write, and work with, their models in terms of various matter fields, each in their separate irreducible representation of the model gauge group. On the other hand, the generic Lagrangian employed in the general result is framed in terms of but one real scalar and one Weyl fermion multiplet in reducible representations of the symmetry, encompassing all the matter fields of the model. Matching the specific model onto the general theory is in principle straightforward; however, in practice, it almost always turns into an arduous exercise and is the primary obstacle in applying the general formulas. 
Typical computations in the generic framework would be done by explicit construction of the coupling tensors, which are mostly large and sparse. In the SM for instance, the generalized Yukawa coupling $ Y_{aij} $ is a sparse $ 4\times 45 \times 45 $ tensor and the fermion gauge generator $ (T^{A}_\Psi)\ud{i}{j} $ an $ 11 \times 45 \times 45 $ tensor. Every term in the \befs corresponds to a monomial in these tensors, contracting all internal indices, and computations quickly become computationally expensive and/or cumbersome to set up. 

For all of the above reasons, a number of computer tools have been developed for extracting model-specific \befs, of which \rgbeta is but the latest. 
There is the general-purpose Mathematica\footnote{Mathematica is a product of Wolfram Research, Inc.} package \texttt{SARAH 4}~\cite{Staub:2008uz,Staub:2013tta}, which includes a 2-loop implementation of the \befs based on the original 2-loop formulas~\cite{Machacek:1983fi,Luo:2002ti}. Then there is the dedicated python code \texttt{PyR@TE 3}~\cite{Sartore:2020gou,Lyonnet:2013dna,Lyonnet:2016xiz}, which in its latest iteration, based on the completely general set of basis tensors~\cite{Poole:2019kcm,Sartore:2020pkk} up to the 3-loop order in the gauge \bef, is blazing fast. Finally, during the completion of this manuscript, we became aware of the independent, dedicated Mathematica package \texttt{ARGES}~\cite{Litim:2020jvl} with a significant overlap with the scope of \rgbeta (even as it pertains to dedicated computation tools for RG functions) and a good degree of flexibility in the implementation of the gauge groups. Also recently, the C++ library \texttt{RGE++} was introduced to provide an easy-to-use framework for numerically solving RG equations in an efficient manner~\cite{Deppisch:2020aoj}.  
Although most of the new features of \rgbeta has been covered in the latest \texttt{PyR@TE} release and \texttt{ARGES}, we still believe it worthwhile to release this latest tool in the line. With multiple available tools, the user can ultimately choose whichever fits them and their project best. Furthermore, with multiple independent codes, it becomes possible to cross-check the outputs. This is sorely needed considering that in all cases reliability of the output is subject to the user not making any errors in the (often complicated) input.
There is also the black-box factor to consider, that is, the implementation in the various programs cannot easily be tested by the users.

\rgbeta was originally developed to test the general \bef basis used and developed in Ref.~\cite{Poole:2019kcm} and is used in ongoing work on higher-order \befs~\cite{Herren}. It is an implementation of the generally applicable \msbar \bef formulas for gauge-Yukawa theories presented in Ref.~\cite{Poole:2019kcm}, which in turn is an extension of the general 2-loop results~\cite{Machacek:1983tz,*Machacek:1983fi,*Machacek:1984zw,Jack:1984vj,Luo:2002ti} and 3-loop gauge \bef~\cite{Pickering:2001aq} to allow for completely generic gauge groups. The new formulas use a revised basis, which easily generalizes to the case with multiple Abelian gauge group factors and has been checked for inconsistencies with the Weyl consistency conditions (cf. Refs.~\cite{Jack:2013sha,Antipin:2013sga,Jack:2014pua}).
\rgbeta uses an implicit construction of the general coupling tensors to avoid large sparse tensors in the evaluation, and all group index contractions are evaluated with a handful of simplification rules and Fierz identities. Although this limits the scope of the package to fields in some of the more common representations of the ordinary Lie groups (cf. Sec.~\ref{sec:group_theory}), it does come with the benefit of a usually quick evaluation time and the possibility of leaving group indices unspecified, meaning that it can handle e.g. $ \SU(N) $ groups without fixing $ N $. This makes \rgbeta ideal for using in the Mathematica notebook environment, where it is easy for the user to manipulate the output and/or make changes to the model on the fly. 

The rest of the paper is organized as follows: The following section details the matching procedure from a specific model to the general framework. It provides background and insight into the underlying procedure but can be skipped by the impatient reader, who wants to learn how to use \rgbeta. Section~\ref{sec:overview} explains the basic objects of the package, installation, and implementation of group theory. Next, Section~\ref{sec:using_RGBeta} outlines the routines of \rgbeta and how to implement a model, using the Standard Model (SM) as a detailed example. Finally, we round off with a short conclusion while the appendix contains more full-fledged documentation of the routines.

\section{Formalism}

\subsection{Generic four-dimensional renormalizable theories}
The derivation of the general formulas for \befs presupposes a certain generic formulation of the QFT. To include all marginal couplings of any four-dimensional theory, the generic Lagrangian (GL) used in the formulation of the general \bef results is compactly put on the form~\cite{Poole:2019kcm}
	\begin{equation}\label{eq:generic_Lagrangian}
	\begin{split}
	\L =\, &-\tfrac{1}{4} a^{\eminus 1}_{AB} F^{A}_{\mu\nu} F^{B\mu\nu} + \tfrac{1}{2} (D_\mu \Phi)_{a} (D^{\mu} \Phi)_a + i \Psi_i^{\dagger} \bar{\sigma}^\mu (D_\mu \Psi)^i \\
	&- \tfrac{1}{2} \left(Y_{aij} \Psi^i \Psi^j \hc \right) \Phi_a - \tfrac{1}{24} \Lambda_{abcd} \Phi_a \Phi_b \Phi_c \Phi_d.
	\end{split}
	\end{equation}
The Fermion field $ \Psi^{i} $ is of the Weyl type, as any Dirac spinor can be decomposed into two 2-component spinors. All fermion fields are gathered into this one multiplet in a, generally speaking, reducible representation of the flavor and gauge group, and lower-case Latin indices starting from $ i,j,\ldots $ are used for the fermion indices.
The scalar field $ \Phi_a $ is taken to be real: complex fields can be decomposed to real fields but not vice versa. $ \Phi_a $, labeled with small Latin letters beginning with $ a,b, \ldots $, is one multiplet containing all scalar fields and is also generically a reducible representation of the gauge and flavor groups. 
Finally, all gauge fields are gathered into a multiplet $ A^{A}_\mu $ with upper-case Latin indices $ A, B,\ldots $ running over all adjoint representations of the individual product groups. $ F^{A}_{\mu\nu} $ is the field-strength tensor associated with the corresponding multiplet. 

As for the couplings, the Yukawa coupling $ Y_{aij} $ is symmetric in the two fermion indices $ i $ and $ j $ and the quartic coupling $ \Lambda_{abcd} $ is completely symmetric. The gauge couplings have been absorbed into the gauge kinetic term and placed in the coupling matrix $ a_{AB}^{\eminus 1} $, which is proportional to the identity in the case of a simple gauge group and otherwise block diagonal (except for kinetic mixing terms between Abelian groups).\footnote{Further details on this construction can be found in Ref.~\cite{Poole:2019kcm}.} Thus, the covariant derivatives of the matter fields read
	\begin{equation} 
	D_\mu \Phi_a = \partial_\mu \Phi_a - i A^{A}_\mu (T_\Phi^{A})_{ab} \Phi_b \andeq D_\mu \Psi^{i} = \partial_\mu \Psi^{i} - i A^{A}_\mu (T_\Psi^{A})\ud{i}{j} \Psi^{j}.
	\end{equation} 

In contrast to the GL, the model builder's Lagrangian (MBL), as it is commonly used in model building, is written in terms of multiple fields in irreducible representations of the gauge and global symmetry groups of the model. It takes the form
	\begin{multline}\label{eq:model_Lagrangian}
	\L = -\tfrac{1}{4} \sum_{n} F_{n,\mu\nu}^{\mathcal{U}_n} F^{\mathcal{U}_u \mu\nu}_{n} 
	+ \sum_\alpha (D_\mu \phi_\alpha)_{\mathcal{A}_\alpha}^\dagger (D^{\mu} \phi_\alpha)^{\mathcal{A}_\alpha} 
	+ i \sum_\rho (\psi_\rho^{\dagger})_{\mathcal{I}_\rho} \bar{\sigma}^\mu (D_\mu \psi_\rho)^{\mathcal{I}_\rho} \\
	-\!\! \sum_{\mathrm{couplings}} \! \left(y_{\mathcal{A}_\alpha \mathcal{I}_\rho \mathcal{J}_\sigma} \phi_\alpha^{\mathcal{A}_\alpha} \psi_\rho^{\mathcal{I_\rho}} \psi_\sigma^{\mathcal{J}_\sigma} \hc \right) 
	- \!\! \sum_{\mathrm{couplings}} \! \lambda_{\mathcal{A}_\alpha \mathcal{B}_\beta \mathcal{C}_\gamma \mathcal{D}_\delta} \phi_\alpha^{\mathcal{A}_\alpha} \phi_\beta^{\mathcal{B}_\beta} \phi_\gamma^{\mathcal{C}_\gamma} \phi_{\delta}^{\mathcal{D}_\delta},
	\end{multline}
where $ n $ labels the various products of the gauge group $ G = \times_n G_n $, and all repeated indices are summed over. This notation is generalized and---unfortunately---rather incomprehensible at a glance.\footnote{No doubt this is what we get for recklessly constructing a generic specific theory: a contradiction in terms.} The Greek subscripts on the fields label the various irreducible representations of the scalar and fermion fields. Each field has a set of indices collectively denoted with a calligraphic letter, which comprises all gauge and flavor indices of the respective fields. 
As an example, the SM left-handed quark doublet would be written as $ \psi_q^{\mathcal{I}_q} $ with the collective index $ \mathcal{I}_q = (i, c, \alpha) $ labeling generation, color, and isospin. For the gauge fields, $ A_n^{\mathcal{U}_n \mu} $ is the field of product group $ G_n $ with corresponding adjoint index $ \mathcal{U}_n $. To apply known results and extract the \befs for the coupling in the MBL, one must construct a mapping between it and the GL.

\subsection{Mapping onto the generic Lagrangian}
Let us proceed with the construction of a mapping of the MBL onto the GL. One approach would be to do an explicit construction of the $ \Phi_a $ and $ \Psi^i $ multiplets with each entry being filled with a particular gauge and flavor component of the fields $ \phi_\alpha $ and $ \psi_\rho $.\footnote{See e.g. App. D of Ref.~\cite{Mihaila:2012pz} for how this would be done in the SM. The reader will no doubt appreciate the effort underlying such explicit constructions and why one might wish to avoid this.} Instead, we will employ and expand on the use of the structure deltas introduced by \citet{Molgaard:2014hpa} in an approach that allows for treating most sums (matrix multiplications) of the program in a mostly implicit manner. 

The peculiar form of the gauge kinetic term in the GL~\eqref{eq:generic_Lagrangian} is a compact way of including all the product groups of a generic semi-simple gauge group $ G = \times_n G_n $. The gauge couplings have been placed in the kinetic term of the gauge fields by rescaling the gauge fields $ A^{\mathcal{U}_n}_{n\mu} \to g_n^{\eminus 1} A^{\mathcal{U}_n}_{n\mu} $. Next, all the gauge fields are collected into a single multiplet $ A^{A}_\mu $. The structure deltas $ \Delta[A_n] $ are introduced as projection operators satisfying
	\begin{equation}
	A_{n\mu}^{\mathcal{U}_n} = \Delta[A_n]^{\mathcal{U}_n A} A^{A}_\mu \andeq
	A^{A}_\mu = \sum_{n} A_{n\mu}^{\mathcal{U}_n} \Delta[A_n]^{\mathcal{U}_n A}.
	\end{equation} 
As projection operators they fulfill the completeness relations 
	\begin{equation}
	\Delta[A_m]^{\mathcal{U}_m A} \Delta[A_n]^{\mathcal{V}_n A} = \delta_{mn} \delta^{\mathcal{U}_m \mathcal{V}_n} \andeq \sum_n \Delta[A_n]^{\mathcal{U}_n A} \Delta[A_n]^{\mathcal{U}_n B} = \delta^{AB}.
	\end{equation}
Using the structure delta, one may write the gauge kinetic term 
	\begin{equation}
	-\tfrac{1}{4} a^{\eminus 1}_{AB} F^{A}_{\mu\nu} F^{B\mu\nu} = -\tfrac{1}{4} \sum_u g_u^{\eminus 2} F_{u,\mu\nu}^{\mathcal{U}_u} F^{\mathcal{U}_u \mu\nu}_{u},
	\end{equation}
by identifying
	\begin{align} \label{eq:g_mapping}
	a_{AB}^{\eminus 1} &= \sum_{n} g_n^{\eminus 2} \Delta[A_n]_{A \mathcal{U}_n } \Delta[A_n]_{B \mathcal{V}_n } \delta_{\mathcal{U}_n \mathcal{V}_n}, \\
	F_{ABC} &= \sum_{n} \Delta[A_n]_{A \mathcal{U}_n } \Delta[A_n]_{B \mathcal{V}_n } \Delta[A_n]_{C \mathcal{W}_n } f_n^{\mathcal{U}_n \mathcal{V}_n \mathcal{W}_n}.
	\end{align}
for the gauge couplings and group structure constants, respectively.  
One may think of $ a^{\eminus 1}_{AB} $ as a block diagonal matrix, where gauge invariance ensures that the value of the coupling is the same for all the gauge fields across the adjoint representation of any one of the product groups $ G_n $. The gauge self-interactions always depend on the generalized structure constant $ F_{ABC} $, which also vanish between gauge fields of different product groups. 

Moving on to the interactions with matter fields, the structure deltas are implemented with the purpose of projecting out the specific fields (in the phenomenological sense) $ \phi_\alpha $ or $ \psi_\rho $ from the general multiplets $ \Phi $ or $ \Psi $, respectively. We define them by
	\begin{align}
	\phi_\alpha^{\mathcal{A}_\alpha} &= \dfrac{1}{\sqrt{2}} \Delta[\phi_\alpha] \ud{\mathcal{A}_\alpha}{a} \, \Phi_a, & 
	\Phi_a &= \dfrac{1}{\sqrt{2}} \sum_\alpha \left( \Delta[\phi_\alpha^\dagger]_{a\mathcal{A}_\alpha} \phi_\alpha^{\mathcal{A}_\alpha} + \Delta[\phi] \ud{\mathcal{A}_\alpha}{a} \phi^\ast_{\alpha,\mathcal{A}_\alpha} \right),
	\\ \psi_\rho^{\mathcal{I}_\rho} &=  \Delta[\psi_\rho]\ud{\mathcal{I_\rho}}{i} \,\Psi^{i},
	& \Psi^{i} &= \sum_\rho \Delta[\psi_\rho^{\dagger}] \ud{i}{\mathcal{I}_\rho} \,\psi_\rho^{\mathcal{I}_\rho},
	\end{align}
where e.g. $ (\Delta[\phi_\alpha]^{\dagger})_{a \mathcal{A}_\alpha} = \Delta[\phi_\alpha^{\dagger}]_{a \mathcal{A}_\alpha} $. The $ \sqrt{2} $ accounts for the different normalization of the kinetic terms of the real and complex scalar fields. Since no degrees of freedom are lost or gained from packing the fields into multiplets, the structure deltas obey the usual summation rules for projection matrices:
	\begin{align} \label{eq:delta_sum}
	\Delta[\phi_\alpha]\ud{\mathcal{A}_\alpha}{a} \Delta[\phi_\beta^{\dagger}]_{a \mathcal{B}_\beta } &= 2 \delta_{\alpha \beta} \delta \ud{\mathcal{A}_\alpha }{\mathcal{B}_\alpha}, 
	&& \sum_\alpha \left(\Delta[\phi_\alpha^{\dagger}]_{a \mathcal{A}_\alpha} \Delta[\phi_\alpha] \ud{\mathcal{A}_\alpha}{b} + \Delta[\phi_\alpha^{\dagger}]_{b \mathcal{A}_\alpha} \Delta[\phi_\alpha] \ud{\mathcal{A}_\alpha}{a}\right)  = 2 \delta_{ab}, \nonumber
	\\ \Delta[\psi_\rho] \ud{\mathcal{I}_\rho}{i} \Delta[\psi_\sigma^{\dagger}] \ud{i}{\mathcal{J}_\sigma} &= \delta_{\rho \sigma} \delta \ud{\mathcal{I}_\rho}{\mathcal{J}_\rho}, 
	&& \sum_\rho \Delta[\psi_\rho^{\dagger}] \ud{i}{\mathcal{I}_\alpha} \Delta[\psi_\rho] \ud{\mathcal{I}_\alpha}{j} = \delta\ud{i}{j}.
	\end{align}
As one would expect, the complex scalars satisfy\footnote{The same applies for fermion fields, but such contractions do not occur in the formalism in the first place.}
	\begin{equation}
	\Delta[\phi_\alpha]\ud{\mathcal{A}_\alpha}{a} \Delta[\phi_\beta]\ud{\mathcal{B}_\beta}{a} = 0 .
	\end{equation}

Using the structure deltas, the interaction terms of the MBL can be recast as 
	\begin{equation}
	\begin{split}
	\L \supset - \dfrac{1}{\sqrt{2}} &\sum_\mathrm{couplings} \left(y_{\mathcal{A}_\alpha \mathcal{I}_\rho \mathcal{J}_\sigma} \Delta[\phi_\alpha] \ud{\mathcal{A}_\alpha}{a} \Delta[\psi_\rho] \ud{\mathcal{I}_\rho}{i} \Delta[\psi_\sigma] \ud{\mathcal{J}_\sigma}{j} \Psi^i \Psi^j \hc \right) \Phi_a \\
	- \dfrac{1}{4} &\sum_{\mathrm{couplings}} \lambda_{\mathcal{A}_\alpha \mathcal{B}_\beta \mathcal{C}_\gamma \mathcal{D}_\delta} \Delta[\phi_\alpha] \ud{\mathcal{A}_\alpha}{a} \Delta[\phi_\beta] \ud{\mathcal{B}_\beta}{b} \Delta[\phi_\gamma] \ud{\mathcal{C}_\gamma}{c} \Delta[\phi_\delta] \ud{\mathcal{D}_\delta}{d} \Phi_a \Phi_b \Phi_c \Phi_d.
	\end{split}
	\end{equation}
Through direct comparison with the GL~\eqref{eq:generic_Lagrangian}, one then finds 
	\begin{align}
	Y_{aij} &= \dfrac{1}{\sqrt{2}} \sum_{\braces{i,j}} \sum_\mathrm{couplings} y_{\mathcal{A}_\alpha \mathcal{I}_\rho \mathcal{J}_\sigma} \Delta[\phi_\alpha]\ud{\mathcal{A}_\alpha}{a} \Delta[\psi_\rho] \ud{\mathcal{I}_\rho}{i} \Delta[\psi_\sigma] \ud{\mathcal{J}_\sigma}{j}, \label{eq:yuk_mapping}
	\\ \Lambda_{abcd} &= \dfrac{1}{4} \sum_{\braces{a, b, c, d}} \sum_\mathrm{couplings} \lambda_{\mathcal{A}_\alpha \mathcal{B}_\beta \mathcal{C}_\gamma \mathcal{D}_\delta} \Delta[\phi_\alpha] \ud{\mathcal{A}_\alpha}{a} \Delta[\phi_\beta] \ud{\mathcal{B}_\beta}{b} \Delta[\phi_\gamma] \ud{\mathcal{C}_\gamma}{c} \Delta[\phi_\delta] \ud{\mathcal{D}_\delta}{d}. \label{eq:lam_mapping}
	\end{align}
The sums over $ \braces{i,j} $ and $ \braces{a, b, c, d} $ are taken to be over all 2 and 24 permutations of the indices, respectively. In a similar manner, the kinetic terms can be made to match the GL form by inserting an identity with a pair of structure deltas. For the fermions it follows that
	\begin{equation}
	\sum_{\rho} \Delta[\psi_\rho^\dagger] \ud{i}{\mathcal{I}_\rho} (D_\mu \psi_\rho )^{\mathcal{I}_\rho} = \bigg( \delta\ud{i}{j} \partial_\mu -i A_\mu^{A} \sum_\rho \Delta[\psi_\rho^\dagger] \ud{i}{\mathcal{I}_\rho} (T^{A}_\rho)\ud{\mathcal{I}_\rho}{\mathcal{J}_\rho} \Delta[\psi_\rho] \ud{\mathcal{J}_\rho}{j} \bigg) \Psi^{j}, 
	\end{equation}
which allows for identifying the gauge generator $ (T^A_\Psi)\ud{i}{j} $ of the fermion field multiplet.  
For the scalars there is a small complication due to the matching from real to complex fields. Observe first that the scalar kinetic term can be put on the form
	\begin{equation}
	\sum_\alpha (D_\mu \phi_\alpha^{\dagger})_\mathcal{A_\alpha} (D^\mu \phi_\alpha)^\mathcal{A_\alpha} =
	\dfrac{1}{4} \sum_{\alpha, \beta} \left[(D_\mu \phi_\alpha^{\dagger})_\mathcal{A_\alpha} \Delta[\phi_\alpha]\ud{\mathcal{A}_\alpha}{a} \hc\right] \left[\Delta[\phi_\beta^{\dagger}]_{a \mathcal{B}_\beta} (D^\mu \phi_\beta)^\mathcal{B_\beta}  \hc\right].
	\end{equation}
Only then do we match to the GL, with 
	\begin{equation}
	\begin{split}
	\tfrac{1}{2} \sum_\alpha &\left[ \Delta[\phi_\alpha^{\dagger}]_{a \mathcal{A}_\alpha} (D_\mu \phi_\alpha)^\mathcal{A_\alpha}  \hc\right] \\
	&= \dfrac{1}{2\sqrt{2}} \sum_\alpha \left( \Delta[\phi_\alpha^{\dagger}]_{a \mathcal{A}_\alpha} D_\mu\ud{\mathcal{A}_\alpha}{\mathcal{B}_\alpha} \Delta[\phi_\alpha]\ud{\mathcal{B}_\alpha}{b} \hc \right) \Phi_b\\
	&= \dfrac{1}{\sqrt{2}} \bigg( \delta_{ab} \partial_\mu - \dfrac{i}{2} A_\mu^{A} \sum_\alpha \left[\Delta[\phi^{\dagger}_\alpha] _{a\mathcal{A}_\alpha} (T^{A}_\alpha) \ud{\mathcal{A}_\alpha}{\mathcal{B}_\alpha} \Delta[\phi_\alpha]\ud{\mathcal{B}_\alpha}{b}  - (a \leftrightarrow b) \right] \! \bigg) \Phi_b.
	\end{split}
	\end{equation}
From here we can identify the generators $ (T^{A}_\Phi)_{ab} $ of the generic scalar multiplet. They are antisymmetric in $ a,b $ as indeed they should be for any real representation.

It should be mentioned that any MBL could, of course, contain real scalars in addition to (or instead of) the complex ones. The formalism presented here is easily extended to accommodate such cases by identifying $ \Delta[\phi_\alpha] $ and $ \Delta[\phi_\alpha^\dagger] $ for the appropriate fields and getting rid of the corresponding $ \sqrt{2} $. 
One great advantage of using the structure deltas is that they never need to be constructed explicitly as matrices. It is sufficient to invoke the completeness relations whenever GL indices are contracted to obtain a sum of contractions with MBL indices.

\subsection{$ \beta $-functions}
Having established a mapping procedure between the MBL and GL, the known results for the Beta functions~\cite{Machacek:1983tz,Machacek:1983fi,Machacek:1984zw,Jack:1984vj,Pickering:2001aq,Luo:2002ti} can be applied to determine the \befs. Specifically, the implementation of the \befs in \rgbeta uses the form given in \cite{Poole:2019kcm}, as it supports completely generic gauge groups including the effects of kinetic mixing up to the 3-loop order. In the GL notation, the general formula for the \befs produces 
	\begin{equation}
	\beta_{AB} = \dfrac{\dd a_{AB}}{\dd t}, \qquad \beta_{aij} = \dfrac{\dd Y_{aij}}{\dd t},  \andeq \beta_{abcd} = \dfrac{\dd \Lambda_{abcd}}{\dd t}, \qquad t=\ln \mu
	\end{equation}
in terms of the couplings $ a_{AB} $, $ Y_{aij} $, and $ \Lambda_{abcd} $ and gauge group generators $ (T_\Phi^A)_{ab} $, $ (T_\Psi^{A})\ud{i}{j} $, and $ F_{ABC} $. As we saw, all of these objects can be expressed with their MBL counterparts and structure deltas, leaving only the task of projecting the generic \befs back onto the specific couplings of the MBL.

The structure deltas are applied to map back from GL \befs to the \befs of the specific MBL couplings. For the gauge couplings we invert Eq.~\eqref{eq:g_mapping}, giving
	\begin{equation}
	\dfrac{\dd g_n^2}{\dd t} = \dfrac{1}{d(G_n)} \beta_{AB} \Delta[A_n]_{A \mathcal{U}_n } \Delta[A_n]_{B \mathcal{U}_n }.
	\end{equation}  
For the Yukawa and quartic interactions, a similar comparison to Eqs.\ \eqref{eq:yuk_mapping} and\ \eqref{eq:lam_mapping} shows that 
	\begin{align}
	\dfrac{\dd y_{\mathcal{A}_\alpha \mathcal{I}_\rho \mathcal{J}_\sigma}}{\dd t} 
	&= \dfrac{1}{\sqrt{2} S_2(\rho, \sigma)} \beta_{aij} \Delta[\phi_\alpha^{\dagger}]_{a \mathcal{A}_\alpha} \Delta[\psi_\rho^{\dagger}]_{i \mathcal{I}_\rho} \Delta[\psi_\sigma^{\dagger}]_{j \mathcal{J}_\sigma}, \label{eq:beta_y}\\
	\dfrac{\dd \lambda_{\mathcal{A}_\alpha \mathcal{B}_\beta \mathcal{C}_\gamma \mathcal{D}_\delta}}{\dd t} 
	&= \dfrac{1}{4 S_4(\alpha, \beta, \gamma, \delta)} \beta_{abcd} \Delta[\phi_\alpha^{\dagger}]_{a \mathcal{A}_\alpha} \Delta[\phi_\beta^{\dagger}]_{a \mathcal{B}_\beta} \Delta[\phi_\gamma^{\dagger}]_{a \mathcal{C}_\gamma} \Delta[\phi_\delta^{\dagger}]_{a \mathcal{D}_\delta}. \label{eq:beta_lam}
	\end{align}
Here we must introduce symmetry factors $ S_n $ to compensate for double counting when multiple fields coincide.\footnote{$ S_n(\alpha, \beta, \ldots) = n! / $the number of unique permutations of the arguments. Thus, e.g. $ S_4(\alpha, \alpha, \beta, \beta) = 4 $.} Contrary to the gauge \bef, both the Yukawa and quartic \befs still have free indices. Some of these are coupling indices, such as the generation indices on the SM Yukawa couplings, and should be retained. However, all free gauge (and flavor indices irrelevant to said coupling) will have to be projected out.  

One should keep in mind a potential problem in projecting out the remaining symmetry indices of Eqs.\ \eqref{eq:beta_y} and\ \eqref{eq:beta_lam}. If the gauge and global symmetry representations of the fields allow for multiple invariants, the naive projection of one coupling may have a non-trivial overlap with the others. In such cases, one must take care to define linear combinations that pick out particular couplings. One example where this becomes relevant is in the single-trace/double-trace terms in quartic potentials.

\subsection{Treatment of other renormalization group functions}
We should like to make some additional comments regarding what RG functions can be extracted with the methods discussed in this section. First of all,  
as pointed out in Ref.~\cite{Poole:2019kcm}, the formulation of the gauge Lagrangian with a matrix gauge coupling generalizes easily to include kinetic mixing terms between multiple Abelian gauge groups, making it possible to treat any gauge group. In contrast to previous formulations of the kinetic-mixing as something other than the gauge coupling~\cite{Luo:2002iq,Fonseca:2013bua}, we can include it by allowing for a matrix gauge coupling between all Abelian field-strength tensors in the gauge potential: $ \L \supset -\dfrac{1}{4} F_{r,\mu\nu} h^{\eminus 1}_{rs} F_s^{\mu\nu} $, where $ r,s $ run over all Abelian gauge fields. This includes all information of the mixing between the groups, and $ h_{rs} $ can be embedded as a block in the general gauge coupling $ a_{AB} $. 

Thus far we have only discussed the \befs of the marginal couplings. Relevant couplings are also allowed in renormalizable theories, and their running is also governed by \befs. The most general Lagrangian term for the relevant couplings take the form	
	\begin{equation}
	\L_\mathrm{rel} = -\tfrac{1}{2} \big(M_{ij} \Psi^i \Psi^j \hc \big) - \tfrac{1}{2} m^2_{ab} \Phi_a \Phi_b - \tfrac{1}{6} h_{abc} \Phi_a \Phi_b \Phi_c,
	\end{equation}
in the GL or   
	\begin{equation}\label{eq:model_Lagrangian_rel}
	\L_\mathrm{rel} = -\sum_{\mathrm{masses}} \left(M_{\mathcal{I}_\rho \mathcal{J}_\sigma} \rho^{\mathcal{I_\rho}} \psi_\sigma^{\mathcal{J}_\sigma} \hc \right) 
	- \sum_{\mathrm{masses}} m^2_{\mathcal{A}_\alpha \mathcal{B}_\beta} \phi_\alpha^{\mathcal{A}_\alpha} \phi_\beta^{\mathcal{B}_\beta} 
	- \!\! \sum_{\mathrm{couplings}} h_{\mathcal{A}_\alpha \mathcal{B}_\beta \mathcal{C}_\gamma} \phi_\alpha^{\mathcal{A}_\alpha} \phi_\beta^{\mathcal{B}_\beta} \phi_\gamma^{\mathcal{C}_\gamma}
	\end{equation}
in the MBL. The structure delta technique can be applied to establish a mapping between the GL and MBL relevant terms in a similar manner to what was just discussed for the marginal couplings. 

The actual \befs for the relevant terms can be recovered from the Yukawa and quartic \befs using the dummy field method~\cite{Schienbein:2018fsw,Luo:2002ti}. It entails introducing non-dynamic faux scalar fields to the relevant couplings to match them with marginal couplings whose \befs can then be used for the relevant couplings. The main pitfall is that as the faux field is not dynamical, it does not receive any field-strength renormalization; all such contributions have to be removed from the corresponding \bef. At the 2-loop order, one can unambiguously identify what terms in the \bef basis corresponds to the (neutral) scalar field anomalous dimensions, and these can therefore be removed without any specific knowledge of the underlying loop calculation. This method was also used by \citet{Sartore:2020pkk} for the new \bef basis~\cite{Poole:2019kcm}.

The anomalous dimension of the matter fields are also known up to the 2-loop order. For a matter field $ \eta $, the bare field is related to the renormalized field by
	\begin{equation}
	\eta_0 = Z_\eta \eta,
	\end{equation}
where $ Z_\eta $ is the field-strength renormalization. The anomalous dimension of the field is then given as 
	\begin{equation} \label{eq:anom_dim}
	\gamma_\eta = Z_\eta^{\eminus 1} \dfrac{\dd}{\dd t} Z_\eta.
	\end{equation}
The general 2-loop expressions found in \citet{Luo:2002ti} are easily adapted to the notation of Ref.~\cite{Poole:2019kcm} used in \rgbeta. It should be pointed out that $ \gamma_\eta $ is gauge dependent, and that they have been computed using the $ R_\xi $ gauges (with $ \xi=0 $ being the Lorenz/Landau gauge and $ \xi=1 $ the Feynman gauge).

\section{Package Overview} \label{sec:overview}

\subsection{RGBeta in a nutshell}
\rgbeta is a Mathematica package aimed at allowing theorists to easily extract the \msbar \befs of their favorite models, be it for BSM physics or field-theoretic applications. The package implements the state-of-the-art \befs for general four-dimensional renormalizable theories, which are known up to loop order 3--2--2, for gauge, Yukawa, and quartic couplings, respectively.
The intent is to provide Mathematica users with an easy-to-use, one-stop implementation of the \bef formulas directly in Mathematica, which is already widely used in the community for many computer-algebra tasks. Specifying a model to the package can be done in a rather compact manner and can easily be done directly in a Mathematica notebook, in which the user can proceed to extract and manipulate the various \befs. The \befs can be obtained modularly for the users who wish to experiment with them in the Mathematica notebook environment.

The core of \rgbeta is set up to evaluate the tensor structures of the general \bef formulas with Einstein summation conventions and various group identities. This approach is similar to what a person might do and is effective because many of the index contractions are in fact Kronecker deltas or group generators, which can be contracted with the pattern matching functionality of Mathematica. The performance of this strategy is generally very good, evaluating e.g. the full set of SM \befs to highest loop order in less than 10 seconds on a decent laptop.\footnote{Other models with matter in the fundamental representations are similarly fast, though needless to say that evaluation time increases with the complexity of the model. The main obstacle to a fast evaluation is when the scalar quartic sector becomes complicated, either because of there being a large number of quartic couplings in the model or because of scalars in non-fundamental representations.} On top of that, it allows the user to keep the number of colors or generations arbitrary during the evaluation. The price of the implementation strategy is that it does require hard-coding various representation-specific identities. For this reason, only a selection of common representations have presently been implemented; cf. Section~\ref{sec:group_theory}.

\subsection{Definition of the \befs}
It pays to be explicit about the definition of the \befs used in \rgbeta, as many different normalizations are used throughout the literature. To reiterate, the package exclusively provides the \msbar (or equivalently MS) \befs. In all cases we associate a loop expansion to the \befs, writing 
	\begin{equation}
	\beta_g = \sum_\ell \dfrac{\beta_g^{(\ell)}}{(4\pi)^{2\ell}}.
	\end{equation}
For all non-gauge couplings, both relevant and marginal, the \befs are defined as the logarithmic derivative of the coupling wrt. the renormalization scale: 
	\begin{equation} \label{eq:beta_yuk_quart}
	\beta_g = \dfrac{\dd}{\dd \ln\mu} \begin{cases}
	y_{aij} & \text{for } g = y\in \braces{\mathrm{Yukawas}} \\
	\lambda_{abcd} & \text{for } g = \lambda \in \braces{\mathrm{quartics}}
	\end{cases}
	\end{equation}
and 
	\begin{equation}
	\beta_g = \dfrac{\dd}{\dd \ln\mu} \begin{cases}
	M_{ij} & \text{for } g = M\in \braces{\text{fermion masses}} \\
	m^2_{ab} & \text{for } g = m^2\in \braces{\text{scalar masses}} \\
	h_{abc} & \text{for } g = h \in \braces{\mathrm{trilinears}}
	\end{cases},
	\end{equation}
where the ordering of flavor indices corresponds to the ordering of fields in the associated coupling.
 In all cases, the coupling will be the reference name of the \bef used in the package. 

The only deviation from this pattern of \bef definitions is the gauge couplings. We will assume that the kinetic terms of the gauge fields and their \befs are normalized as 
	\begin{equation}
	\L \supset \sum_n -\dfrac{1}{4 g_n^2} (F_n^{\mathcal{U}_n,\mu\nu})^2, \qquad \beta_{g_n} = \dfrac{\dd g_n^2}{\dd \ln \mu}
	\end{equation}
for the fields of non-Abelian gauge groups or the Abelian gauge group in a model with \emph{at most one} Abelian group. For a model with multiple Abelian gauge groups, which therefore features kinetic mixing, the kinetic term for all the Abelian gauge fields is written jointly as
	\begin{equation} \label{eq:beta_abelian}
	\L \supset -\dfrac{1}{4} F_{r,\mu\nu} h^{\eminus 1}_{rs} F_s^{\mu\nu}, \qquad \beta_h = \dfrac{\dd h_{rs} }{\dd \ln \mu},
	\end{equation}  
where $ h_{ij} $ is the symmetric coupling matrix.\footnote{For further details on how this compares to previous implementations of the kinetic mixing, we refer to the appendix of Ref.~\cite{Poole:2019kcm}.}
The \bef $ \beta_h $ of the coupling matrix contains all information of the running of couplings and mixing.  
With the normalization of the gauge fields employed in this notation, the covariant derivative of a matter field can be written as 
	\begin{equation}
	D_\mu \eta = \partial_\mu \eta - i \sum_n A_{n,\mu}^{\mathcal{U}_n} T^{\mathcal{U}_n}_n \eta - i \sum_r q_r A_{r,\mu} \eta,  
	\end{equation}
where $ T^{A}_n $ are Hermitian generators of the non-Abelian representations and $ q_i $ Abelian charges. In particular, using a coupling matrix for the Abelian kinetic term allows for keeping the coupling of the Abelian gauge fields to the matter fields \quote{diagonal}.

The matter field anomalous dimensions are loop expanded as well, parameterizing
	\begin{equation}
	\gamma_\eta = \sum_\ell \dfrac{\gamma_\eta^{(\ell)}}{(4\pi)^{2\ell}} \, .
	\end{equation}
They are defined in Eq.~\eqref{eq:anom_dim}, and the indices in \rgbeta are
	\begin{equation}
	\gamma_\eta = \begin{dcases}
	\gamma\ud{i}{j} & \text{for } \eta \in \braces{\text{fermions}} \\ 
	\gamma_{ab} & \text{for } \eta \in \braces{\text{scalars}}
	\end{dcases},
	\end{equation} 
all calculated in the $ R_\xi $ gauges.

\subsection{Installation}
\rgbeta can be installed directly to the \path{/Applications} folder in Mathematica's \lstinline|$UserBaseDirectoy|by running the command 
\begin{lstlisting}[numbers=none]
Import["https://raw.githubusercontent.com/aethomsen/RGBeta/master/Install.m"]
\end{lstlisting}
in a Mathematica notebook. Once installed, the package can be loaded into the kernel of any notebook with 
\begin{lstlisting}[numbers=none, mathescape =true]
<< RGBeta$ \grave{} $
\end{lstlisting}
As an alternative to a true installation, \rgbeta can be downloaded from the \texttt{GitHub} repository \href{https://github.com/aethomsen/RGBeta}{github.com/aethomsen/RGBeta \faicon{github}}.
It can then be loaded by simple plug-and-play: put the package in any directory, point a Mathematica notebook in the right direction, and load the package. Most simply, \rgbeta can be loaded from a Mathematica notebook in the base directory of the package (the one with \texttt{README.md}) with 
\begin{lstlisting}[numbers=none, mathescape =true]
SetDirectory@ NotebookDirectory[ ];
<< RGBeta$ \grave{} $
\end{lstlisting}
In all cases, \rgbeta ought to be loaded into a freshly initialized Mathematica kernel to avoid clashing symbol definitions.

\subsection{Group theory} \label{sec:group_theory}
\rgbeta is set up to evaluate the tensor structures in the \bef formulas 
in a manner similar to what a person would do if asked to use the 1-loop formulas by hand for a particular model (something which is typically doable with a bit of patience). That is to say, it avoids all explicit summation/matrix multiplication, except over the various field types, in favor of relying on implicit summation with Einstein summation convention for gauge and flavor indices. This is possible, and fast, for several of the common representations for the ordinary Lie groups, where gauge indices are typically contracted with Kronecker deltas or generators of the fundamental representations, which can be dealt with using Fierz identities. This approach has the benefit that it allows for keeping e.g. the index of the gauge group unspecified throughout the evaluation.    

\begin{table}
	\centering
	\begin{tabular}{|l| p{.7\textwidth}|}
		\hline \hline
		Symbol & 
		Interpretation\\ \hline
		\lstinline[emph={x}]|Bar[x]| 
		& Complex conjugation, $ x^\ast $. It is used general purpose for both fields, couplings, and group representations. 
		\\ \lstinline[emph={x}]|Trans[x]| 
		& Symbolic transposition of matrix couplings.
		\\ \hline \lstinline[emph={rep, a, b}]|del[rep, a, b]| 
		& The Kronecker delta $ \delta_{ab} $ with indices running in the group representation or over flavor indices as specified by \argument{rep}.
		\\ \lstinline[emph={group, a, b}]|delA2[group, i, a, b]| 
		& The Clebsch--Gordan coefficient between indices $ i $ in $ \repbar{A}_2 $ and $ a,b $ in $ \rep{N} $ (or $ i $ in $ \rep{A}_2 $ and $ a,b $ in $ \repbar{N} $) representations of \lstinline[emph=group]|group|.
		\\ \lstinline[emph={rep, a, b}]|delIndex[rep, a, b]| 
		& The Kronecker delta $ \delta_{ab} $, where one of $ a, b $ is an integer pointing to a definite index value, e.g. $ \delta_{a2} $. 
		\\ \lstinline[emph={group, a, b}]|delS2[group, i, a, b]| 
		& The Clebsch--Gordan coefficient between indices $ i $ in $ \repbar{S}_2 $ and $ a,b $ in $ \rep{N} $ (or $ i $ in $ \rep{S}_2 $ and $ a,b $ in $ \repbar{N} $) representations of \lstinline[emph=group]|group|.
		\\ \lstinline[emph={rep, a, b}]|eps[rep, a, b]| 
		& The 2-index antisymmetric invariant $ \epsilon_{ab} $ of a pseudoreal representation, \argument{rep}, such as the fundamental of an $ \Sp(N) $ group. 
		\\ \lstinline[emph={rep, a, b},mathescape]|lcSymb[rep, a, b,$ \ldots $]| & The Levi-Civita symbol with indices in the given representation.  
		\\ \lstinline[emph={rep, A, a, b}]|tGen[rep, A, a, b]|.
		& The Hermitian group generator $ T\ud{Aa}{b} $ for the corresponding representation. The first index, $ A $, is always taken to be in the adjoint representation, while the other two belong to \lstinline|rep|. 
		\\ \lstinline[emph={rep}]|Dim[rep]| 
		& The dimension of a representation or flavor index. The gauge representations are mostly predefined, but the dimension of flavor indices can be set by the user.
		\\ \hline \lstinline[emph={A, B, a, b}, mathescape = true]|Matrix[A, B,$ \ldots $][a, b]| 
		& The matrix product of multiple couplings with 2 or fewer indices, e.g. $ (A B)_{ab} = A_{ac} B_{cb} $. Couplings with 1 index are interpreted as column vectors. 
		\\ \lstinline[emph={A, a, }, mathescape = true]|Tensor[A][a,$ \ldots $]| & The tensor coupling, $ A $, with three or more indices. No coupling contractions involving tensors are supported. 
		\\ \hline \hline
	\end{tabular} 
	\caption{List of various symbols used in \rgbeta and their meaning.}
	\label{tab:symbols}
\end{table}

Using implicit summation, \rgbeta is set up with a number of symbols with internal summation rules set as Mathematica up-values, a list of which is presented in Table~\ref{tab:symbols}. The most basic symbol used in the package is the Kronecker delta $ \delta_{ab} $, which is represented by \lstinline[emph={rep, a, b}]|del[rep, a, b]| with the first argument specifying the type of the indices $ a,b $. The summation conventions assigned to the Kronecker delta makes contractions evaluate as e.g. 
\begin{example}
In[1]		del[rep, a, b] del[rep, c, b] del[rep, c, a]
Out[1]		Dim[rep]
\end{example}
for any representation \lstinline|rep|. Identically named indices of different types do not contract, so e.g. 
\begin{example}
In[2]		del[rep1, a, b] del[rep2, a, b] del[rep2, c, b] del[rep1, b, a]
Out[2]		del[rep2,a,c] Dim[rep1]
\end{example}
allowing for reusing index names. 

\begin{table} \renewcommand*{\arraystretch}{2.5}
	\centering 
	\begin{tabular}{|l r| c c c | c|} \hline \hline
	\multicolumn{2}{|l|}{Representation} & $ d(R) $ & $ S_2(R) $ & $ C_2(R) $ & \rgbeta reference \vspace{-.5em} \\ \hline
	$ \SU(N) $ & $ \mathbf{N} $ & $ N $ & $ \dfrac{1}{2} $ & $ \dfrac{N^2-1}{2N} $ & \lstinline|Gr[fund]| \\
	& $ \mathbf{G} $ & $ N^2 -1 $ & $ N $ & $ N $ & \lstinline|Gr[adj]| \\
	& $ \mathbf{S}_2 $ & $ \dfrac{N(N+1)}{2} $ & $ \dfrac{N+2}{2} $ & $ \dfrac{(N-1) (N+2)}{N} $ & \lstinline|Gr[S2]|\\
	& $ \mathbf{A}_2 $ & $ \dfrac{(N-1) N}{2} $ & $ \dfrac{N-2}{2} $ & $ \dfrac{(N-2) (N+1)}{N} $ & \lstinline|Gr[A2]| \\ \hline
	$ \SO(N) $ & $ \mathbf{N} $ & $ N $ & $ \dfrac{1}{2} $ & $ \dfrac{(N-1)}{4} $ & \lstinline|Gr[fund]| \\
	& $ \mathbf{G} $ & $ \dfrac{(N -1) N}{2} $ & $ \dfrac{N-2}{2} $ & $ \dfrac{N-2}{2} $ & \lstinline|Gr[adj]| \\
	& $ \mathbf{S}_2 $ & $ \dfrac{(N-1)(N+2)}{2} $ & $ \dfrac{N+2}{2} $ & $ \dfrac{N}{2} $ & \lstinline|Gr[S2]|\\ \hline
	$ \Sp(N) $ & $ \mathbf{N} $ & $ N $ & $ \dfrac{1}{2} $ & $ \dfrac{(N+1)}{4} $ & \lstinline|Gr[fund]| \\
	& $ \mathbf{G} $ & $ \dfrac{N (N+1)}{2} $ & $ \dfrac{N+2}{2} $ & $ \dfrac{N+2}{2} $ & \lstinline|Gr[adj]| \\
	& $ \mathbf{A}_2 $ & $ \dfrac{(N-2)(N+1)}{2} $ & $ \dfrac{N-2}{2} $ & $ \dfrac{N}{2} $ & \lstinline|Gr[A2]|\\ \hline
	$ \U(1) $ & $ q $ & 1 & $ q^2 $ & $ q^2 $ & \lstinline[emph={q}]|Gr[q]| \\
	$ \U^n(1) $ & $ (q_1, \ldots, q_n) $ & 1 & --- & --- & \lstinline[emph={q1, q2}]|Gr[{q1, q2,...}]| \\\hline \hline
	\end{tabular}
	\caption{Group constants for the Lie Groups representations implemented in \rgbeta. The group representations are all referred to by \lstinline[]|Gr[rep]| in \rgbeta, where \lstinline|Gr| is the user-chosen name for the group of the specific type. Conjugate representations are referred to as \lstinline[]|Bar[Gr[rep]]|.}
	\label{tab:group_representations}
\end{table}

\begin{table} \renewcommand*{\arraystretch}{2}
	\centering
	\begin{tabular}{|l | c|} \hline \hline
	Group & Fierz identity \vspace{-.4em} \\ \hline
	$ \SU(N) $ & $ T\ud{Ai}{j} T\ud{Ak}{\ell} = \dfrac{1}{2} \delta\ud{i}{\ell} \delta\ud{k}{j} - \dfrac{1}{2N} \delta \ud{i}{j} \delta\ud{k}{\ell} $ \\
	$ \SO(N) $ & $  T^{A}_{ij} \, T^{A}_{k \ell} = \tfrac{1}{4} \left(\delta_{i\ell} \delta_{jk} - \delta_{ik} \delta_{j\ell}\right) $\\
	$ \Sp(N) $ & $ T\ud{Ai}{j} T\ud{Ak}{\ell} = \tfrac{1}{4} \left( \epsilon^{ik} \epsilon_{j \ell} + \delta\ud{i}{\ell} \delta\ud{k}{j} \right) $ \\ \hline \hline
	\end{tabular}
	\caption{Fierz identities for the fundamental representations of the ordinary, compact Lie groups. We employ the convention $ \epsilon_{ij} = \epsilon^{ji} $.}
	\label{tab:Fierz_identities}
\end{table}

The up-value approach is extended to the treatment of gauge generators \lstinline[emph={rep, A, a, b}]|tGen[rep, A, a, b]| with rules for simplifying to group constants. With an irreducible representation \lstinline|G[rep]| of group \lstinline|G|, \rgbeta therefore gives
\begin{example}
In[3]		tGen[G@ rep, A, a, b] tGen[G@ rep, A, b, c]
In[4]		tGen[G@ rep, A, a, b] tGen[G@ rep, B, b, a]
Out[3]		Casimir2[G[rep]] del[G[rep], a, c]
Out[4]		TraceNormalization[G[rep]] del[G[adj], A, B]
\end{example}
Here \lstinline|Casimir2[G[rep]]| is the value of the quadratic Casimir of the representation, and \linebreak \lstinline[breaklines=false]|TraceNormalization[G[rep]]| is the trace normalization/Dynkin index. The values of Casimir operators and Dynkin indices are predefined for all the implemented representations and are available in Table~\ref{tab:group_representations}.

The tensor structures used for the 3--2--2 \befs have been chosen such as to minimize the occurrence of non-trivial group structures stemming from the gauge groups~\cite{Poole:2019kcm}. Despite this, there remains a couple of them that cannot be evaluated by identifying factors of quadratic Casimir operators or Dynkin indices. These tensors are dealt with by implementing the Fierz identities of the fundamental representations of the groups (Table~\ref{tab:Fierz_identities}). In the SM, these are necessary in the 3-loop gauge \befs and the quartic \bef already from 1-loop order. 

The need for Fierz identities and systematic symbolic treatment of group invariants are the main obstacles to implementing arbitrary representations of the gauge groups. At present \rgbeta is, thus, restricted to the representations and groups listed in Table~\ref{tab:group_representations}. In addition to fundamental and adjoint representations, $ \mathbf{N} $ and $ \mathbf{G} $, of the ordinary groups, it includes the two-index symmetric and antisymmetric, $ \mathbf{S}_2 $ and $ \mathbf{A}_2 $, of $ \SU(N) $; the traceless two-index symmetric $ \mathbf{S}_2 $ of $ \SO(N) $; and the two-index antisymmetric that vanishes upon contraction with $ \epsilon_{ab} $, $ \mathbf{A}_2 $, of $ \Sp(N) $. For all these representations, the generators can be decomposed in terms of the generators of the fundamental representation after which the Fierz identities can be employed. The $ \U^n(1) $ groups support representations with any charge assignments. The limit to these few, if frequently used, representations of the Lie groups is the primary limitation of \rgbeta. None of the exceptional Lie groups or their representations are presently implemented.  

The 2-index representations $ \rep{A}_2$ and $ \rep{S}_2 $ are implemented using a single index in \rgbeta (when referring to them as 2-index representation, it is because they can be written in terms of 2 fundamental indices, similar to how an adjoint representation can be written with a fundamental and an anti-fundamental index). We need a way to contract e.g. the $ \rep{A}_2 $ label with two fundamental indices to use as an invariant in various interactions. For this purpose \rgbeta includes the Clebsch--Gordan coefficients \lstinline|delA2| and \lstinline|delS2| (Table~\ref{tab:symbols}). These are really just shortcuts and could easily be constructed using the regular \lstinline|del| by accessing the \quote{subindices} \lstinline|a[1]| and \lstinline|a[2]| of e.g. the $ \rep{A}_2 $ index \lstinline|a|. Thus, \lstinline|delA2[G, i, a, b]| can be reproduced as \lstinline|(del[G[fund], i[1], a] del[G[fund], i[2], b] - del[G[fund], i[2], a] del[G[fund], i[1], b])/2| for any group \lstinline|G| with an $ \rep{A}_2 $ representation (similarly for $ \rep{S}_2 $). It is always preferred to use \lstinline|delA2| and \lstinline|delS2| directly, as they have additional internal identities.

\subsection{Validation of the program}
Like any other computer tool, \rgbeta is for all intents a black box to the user. It does not matter that the underlying theory is sound, if there is a mistake in the implementation, something which can easily go unnoticed. To avoid this kind of errors, we have validated \rgbeta against several results from literature and \texttt{PyR@TE 3}. The \rgbeta implementation of all the models used for validation are included with the package in \path{/Documentation/Sample_Models.nb}.  

First, we compared directly with the SM and type-III 2HDM (two-Higgs-doublet model) computation~\cite{Herren:2017uxn} up to order 3--2--1 with matrix Yukawa couplings. This comparison allowed us to settle a discrepancy in the 2-loop Yukawa \bef between Refs.~\cite{Machacek:1983fi,Luo:2002ti} and \cite{Jack:1984vj,Poole:2019kcm}, settling in favor of Refs.~\cite{Jack:1984vj,Poole:2019kcm} (recall that \rgbeta is based on Ref.~\cite{Poole:2019kcm}).\footnote{The 2-loop Yukawa \befs of Refs.~\cite{Machacek:1983fi,Luo:2002ti} can be corrected by substituting $ -4 \kappa Y^{ac}_2(S) Y^b Y^{\dagger c} Y^b \to -4 \kappa Y^{bc}_2(S) Y^b Y^{\dagger a} Y^c  $. The original term corresponds neither to a 1PI leg or a 1PI vertex contribution, suggesting that it was indeed a typo.} The terms in question are degenerate in the SM but not in the 2HDM. Furthermore, a comparison with the 2HDM result~\cite{Herren:2017uxn} revealed a typo in the 3-loop gauge \befs of their result, which has since been fixed. We also have agreement with the 2-loop SM quartic result of Ref.~\cite{Bednyakov:2013cpa}.

To test the handling of unspecified gauge and flavor groups, we reproduced the 3--2--1 \befs of the Litim-Sannino model, keeping $ N_f $ and $ N_c $ as free variables, and found complete agreement with Ref.~\cite{Litim:2014uca}. The implementation of gauge kinetic mixing was tested in the SM with a gauged $ \U(1)_{B-L} $ symmetry (see e.g.~\cite{Basso:2010jm}) against the \texttt{PyR@TE 3} results~\cite{Sartore:2020gou}. Likewise, we compared the results for the $ \SU(5) $ grand unified theory (GUT).\footnote{The 2-loop $ \SU(5) $ quartic \befs evaluate dreadfully slow (well over an hour on a laptop) in \rgbeta. This is due to the model involving scalars in the adjoint representation, where there are no available Fierz identities. Instead, the generators, where necessary, are expressed as traces over fundamental generators. This is much less of a problem in the 2-loop Yukawa \bef, as the tensors reduce to simple group invariants.} In both cases complete agreement was found up to order 3--2--2.

\section{Using RGBeta} \label{sec:using_RGBeta}
This section describes how to use \rgbeta in some detail. We heartily recommend prospective users to have a look at the tutorial notebook \path{/Documentation/Tutorial.nb}, which illustrates some practical use cases in the form of the SM and Litim-Sannino model~\cite{Litim:2014uca} with comments.

\subsection{Package routines}
To use the \rgbeta package, one should start by specifying a model. First, the user should define symmetries of the model with the routines  
	\begin{itemize}[label={\tiny$\blacksquare$}]
	\item \lstinline[emph={coupling, groupName, lieGroup, n}]|AddGaugeGroup[coupling, groupName, lieGroup[n], Options]| defines a gauge group of type \lstinline[emph={lieGroup, n}]|lieGroup[n]| with reference name \lstinline[emph={groupName}]|groupName| and associated \lstinline[emph={coupling}]|coupling|;
	\item \lstinline[emph={groupName, lieGroup, n}]|DefineLieGroup[groupName, lieGroup[n]]| defines a global symmetry group of type \lstinline[emph={lieGroup, n}]|lieGroup[n]| with reference name \lstinline[emph={groupName}]|groupName|. 
	\end{itemize}
Next, the matter fields are specified with  
	\begin{itemize}[label={\tiny$\blacksquare$}]
	\item \lstinline[emph={field}]|AddFermion[field, Options]| defines a Weyl fermion field;
	\item \lstinline[emph={field}]|AddScalar[field, Options]| defines a scalar field.
	\end{itemize}
At this point one can proceed to add the remaining couplings:
	\begin{itemize}[label={\tiny$\blacksquare$}]
	\item \lstinline[emph={coupling, psi1, psi2}]|AddFermionMass[coupling, {psi1, psi2}, Options]| defines a mass term between two fermion fields, \lstinline[emph=psi1]|psi1| and \lstinline[emph=psi1]|psi2|;
	\item \lstinline[emph={coupling, phi1, phi2, phi3, phi4}]|AddQuartic[coupling, {phi1, phi2, phi3, phi4}, Options]| defines a quartic coupling between four scalar fields;
	\item \lstinline[emph={coupling, pwi1, pwi2}]|AddScalarMass[coupling, {psi1, psi2}, Options]| defines a scalar mass term between two scalar fields;
	\item \lstinline[emph={coupling, phi1, phi2, phi3}]|AddTrilinear[coupling, {phi1, phi2, phi3}, Options]| defines a trilinear coupling between three scalar fields; 
	\item \lstinline[emph={coupling, phi, psi1, psi2}]|AddYukawa[coupling, {phi, psi1, psi2}, Options]| defines a Yukawa coupling between a scalar field, \lstinline[emph=phi]|phi|, and two fermion fields, \lstinline[emph=psi1]|psi1| and \lstinline[emph=psi1]|psi2|.
	\end{itemize}
A few other routines might be helpful at this point:
	\begin{itemize}[label={\tiny$\blacksquare$}]
	\item \lstinline[emph={coupling}]|CheckProjection[coupling]| returns the specific coupling projector applied to the generic vertex of the appropriate type. It is useful for checking if everything is correctly implemented; 
	\item \lstinline|ResetModel[]| clears the kernel of all model definitions. This should be called before implementing a new model in the same session;
	\item \lstinline[emph={symbol}]|SetReal[symbol,...]| sets a symbol, i.e. a coupling, to be treated as being real.
	\end{itemize}
This is all that is necessary for defining the model. Once the model has been loaded into the kernel, the RG functions can be extracted with a minimum of effort using the routines
	\begin{itemize}[label={\tiny$\blacksquare$}] 
	\item \lstinline[emph={field, loop}]|AnomalousDimension[field, loop, Options]| returns the anomalous dimension of the matter \lstinline[emph=field]|field| up to loop order \lstinline[emph=loop]|loop|;
	\item \lstinline[emph={field, loop}]|AnomalousDimTerm[field, loop]| returns the \lstinline[emph=loop]|loop|-loop term of the \lstinline[emph=field]|field| anomalous dimension;
	\item \lstinline[emph={coupling, loop}]|BetaFunction[coupling, loop, Options]| returns the full \bef of the \lstinline[emph=coupling]|coupling| evaluated up to loop order \lstinline[emph=loop]|loop|;
	\item \lstinline[emph={coupling, loop}]|BetaTerm[coupling, loop]| returns the \lstinline[emph=loop]|loop|-loop term of the \lstinline[emph=coupling]|coupling| \bef;
	\item \lstinline[emph={expr}]|Finalize[expr, Options]| returns a refined version of \lstinline[emph={expr}]|expr|, such as \befs or anomalous dimension terms;
	\item \lstinline[emph={loop}]|QuarticBetaFunctions[loop, Options]| returns all the quartic \befs up to the given loop order, fully diagonalizing the coupling projectors. 
	\end{itemize}
The Appendix contains more detailed documentation for the various routines.
Here, we explain the basic use of the package in the next subsections with a detailed example.

\subsection{Setting up a model}
We detail the core use of \rgbeta with the SM as a concrete example, as it will be familiar to most/all users. The package is set to work with Weyl spinors to be able to treat all fermions in the same manner. The SM matter fields have charge assignments 
	\begin{align}
	q & \in (\rep{3},\, \rep{2},\, 1/6), \qquad & \bar{u} &\in (\repbar{3},\, \rep{1},\, 2/3), \qquad & \bar{d} &\in (\repbar{3},\, \rep{1},\, -1/3), \nonumber\\
	\ell & \in (\rep{1},\, \rep{2},\, 1/2), \qquad & \bar{e} &\in (\rep{1},\, \rep{1},\, -1), \qquad & H &\in (\rep{1},\, \rep{2},\, 1/2) 
	\end{align}
under the gauge group $ G_\mathrm{SM} = \SU(3)_c \times \SU(2)_\LL \times \U(1)_Y $. 
There are three Yukawa couplings given by 
	\begin{equation} \label{eq:SM_yukawa}
	\L_\mathrm{yuk} = - y_u^{ij} H_\alpha^{\ast} \epsilon^{\alpha \beta} q\ud{\dagger i}{c \beta} \bar{u}^{\dagger jc} - y_d^{ij} H^\alpha q\ud{\dagger i}{c \alpha} \bar{d}^{\dagger jc} - y_e^{ij} H^\alpha \ell\ud{\dagger i}{\alpha} \bar{e}^{\dagger j} \hc,
	\end{equation}
where $ c $ is a color index, $ \alpha, \beta $ $ \SU(2)_\LL $ indices, and $ i,j $ generation indices. The reason for specifying the Yukawa couplings with right-handed (Hermitian-conjugated) fermions is to match the conventional definition of the couplings in the Dirac notation.\footnote{With Dirac fields, the SM Yukawa couplings are typically written as \begin{equation*} \L_\mathrm{yuk} = - y_u^{ij} H_\alpha^{\ast} \epsilon^{\alpha \beta} \overline{Q}_\LL\ud{i}{c \beta} U\du{\RR}{jc} - y_d^{ij} H^\alpha \overline{Q}_\LL\ud{i}{c \alpha} D\du{\RR}{jc} - y_e^{ij} H^\alpha \overline{L}_\LL \ud{i}{\alpha} E\du{\RR}{j} \hc, \end{equation*}} The SM also contains a scalar potential with a quartic Higgs interaction and a Higgs mass term:
	\begin{equation} \label{eq:SM_quartic}
	\L_V = - M^2 H^\ast_\alpha H^\alpha- \tfrac{1}{2} \lambda (H^\ast_\alpha H^\alpha)^2.
	\end{equation}

The \rgbeta package automates most of the process of getting the \befs. It is, however, unavoidable that the user will have to input the model in a precise manner. Specifying gauge groups and scalar and fermion field content is fairly straightforward. Arguably, the most difficult part in using \rgbeta is specifying the Yukawa and quartic interaction: the user must manually specify how the flavor and gauge indices are contracted, which is a potential source of errors. 

The first thing to do when defining a model is to specify the gauge groups to \rgbeta with \lstinline|AddGaugeGroup|. 
For each product group, one needs simply to specify the coupling, group name, and Lie group. The choice of Lie group will then set up the group invariants and generator properties of the supported representations. The unique name of the group is used for referencing the group representations.
For the SM, the $ \SU(3)_c \times \SU(2)_\LL \times \U(1)_Y $ gauge group is added with 
\begin{lstlisting}[firstnumber = 1]
AddGaugeGroup[g1, U1Y, U1]
AddGaugeGroup[g2, SU2L, SU[2]]
AddGaugeGroup[g3, SU3c, SU[3]]
\end{lstlisting}

Having set up the gauge group, the next step is adding the matter content with \lstinline|AddFermion| and \lstinline|AddScalar|. It is sufficient to know the charges and flavor indices of the fields to do so. As previously mentioned, all fermions must be added as left-handed chiral fields. If any of the fields in the model are given as right-handed fermions, simply add it as a left-handed spinor in the \emph{conjugate} representation under the symmetries. 
The fermions are specified with a field name, and all their \emph{non-trivial} gauge charges are given as a list with the \lstinline|GaugeRep| option (the default being no gauge charges). A full list of the available representations is given in Table~\ref{tab:group_representations}. Given that the fields come in three generation, they are also given a single flavor index with the \lstinline|FlavorIndices| option.\footnote{The flavor group of the SM is really $ \U(3)^5 $, so one can argue that there really are five distinct generation indices. This does not matter for our purposes here, and would only serve as a needless complication.} To add the fermions to the model, we call 
\begin{lstlisting} [firstnumber = 4]
AddFermion[q, GaugeRep->{U1Y[1/6], SU2L[fund], SU3c[fund]}, FlavorIndices->{gen}]
AddFermion[u, GaugeRep->{U1Y[-2/3], Bar@ SU3c[fund]}, FlavorIndices->{gen}]
AddFermion[d, GaugeRep->{U1Y[1/3], Bar@ SU3c[fund]}, FlavorIndices->{gen}]
AddFermion[l, GaugeRep->{U1Y[-1/2], SU2L[fund]}, FlavorIndices->{gen}]
AddFermion[e, GaugeRep->{U1Y[1]}, FlavorIndices->{gen}]
Dim[gen] = ng; (* 3 *)
\end{lstlisting}
The last line may seem a bit curious. It is simply there to specify the dimension of the flavor index \quote{\lstinline|gen|.} In the SM there are, of course, three generations, but it is often kept as a free parameter in \bef computations. Either can be used in \rgbeta. 

We proceed to add the Higgs field in a similar manner. It has just one generation, so there is no need to give it any flavor indices:  
\begin{lstlisting}[firstnumber = 10]
AddScalar[H, GaugeRep->{U1Y[1/2], SU2L[fund]}]
\end{lstlisting}
The package assumes scalar fields to be complex-valued by default, as is the SM Higgs field. If we wished to add a real scalar field, we should give the option \lstinline|SelfConjugate->True|. In models with exact global symmetries, the flavor group can be added with \lstinline|DefineLieGroup|, and the field representations can be specified by using the group representation in the\lstinline|FlavorIndices| option for the fields. 

The most involved part of defining the model is setting up the couplings. In particular, one must take great care when defining the contraction of all field indices and specify the indices of the couplings. The particulars of this should be passed to the package in terms of pure functions. It is \emph{important} to always remember to surround pure functions given as options with parentheses. Otherwise, the function call will not properly evaluate. 
Let us begin by listing the code needed to specify the SM Yukawa couplings before we dissect it:
\begin{lstlisting}[firstnumber=11]
AddYukawa[yu, {H, q, u},
	GroupInvariant->(del[SU3c@ fund, #2, #3] eps[SU2L@ fund, #1, #2] &),
	CouplingIndices->({gen[#2], gen[#3]} &),
	Chirality->Right]
AddYukawa[yd, {Bar@ H, q, d},
	GroupInvariant->(del[SU3c@ fund, #2, #3] del[SU2L@ fund, #1, #2] &),
	CouplingIndices->({gen[#2], gen[#3]} &),
	Chirality->Right]
AddYukawa[ye, {Bar@ H, l, e},
	GroupInvariant->(del[SU2L@ fund, #1, #2] &),
	CouplingIndices->({gen[#2], gen[#3]} &),
	Chirality->Right]
\end{lstlisting}
The first \lstinline|AddYukawa| call specifies the up-type Yukawa coupling. The first two arguments gives the coupling name, \lstinline|yu|, and the fields entering the interaction, \lstinline|{H, q, u}|, the first of which is always taken to be the scalar. Do bear in mind that the order of the fields matter for the construction of the index contractions.  

The \lstinline|GroupInvariant| option should be given a function with three arguments that specify the contraction between the indices of the fields. As the name suggests, this should be an invariant of the gauge and global symmetries. 
It is a function because it will be called every time the coupling appears in the \bef formulas to give appropriate labels to the invariant. 
For the up-type Yukawa, the \lstinline|del[SU3c@  fund, #2, #3]| part of the group invariant establishes that the fundamental $ \SU(3)_c $ indices of the second and third fields are contracted with a Kronecker delta. Furthermore, the  fundamental $ \SU(2)_\LL $ index of the Higgs (the first field) is contracted by an $ \epsilon $-invariant of $ \SU(2)_\LL $ with the left-handed quark (second field) as specified by \lstinline|eps[SU2L@  fund, #1, #2]|.\footnote{The various invariants are multiplied with each other in the function.} 
The remaining free indices of the fields are the generation indices of the quarks. The coupling itself carries these indices, which we spell out with the option \lstinline|CouplingIndices -> ({gen[#2], gen[#3]} &)|. This option should be passed a function of 3 arguments (one for each field) that returns a list of the indices of the coupling. In practice we can almost think of the arguments \lstinline|#1|, \lstinline|#2|, and \lstinline|#3| as the indices of the fields, and it is not a problem to have e.g. multiple \lstinline|#2| indices when they belong to different representations: there is no ambiguity. 

The last option passed to \lstinline|AddYukawa| is more of a quality-of-life option than strictly necessary. The Yukawa couplings in \rgbeta are always given in terms of left-handed spinors as in the MBL~\eqref{eq:model_Lagrangian}, but conventionally the SM defines the Yukawa couplings on the right-handed spinors, and the complex conjugate of the coupling appears with the left-handed fermions. To tell \rgbeta that the un-conjugated coupling should be placed with the right-handed spinors, we need to pass it the option \lstinline|Chirality-> Right| (\lstinline|Left| by default). 
The other two Yukawa couplings follow in mostly the same manner as the up-type Yukawa, the main difference being that they involve the complex conjugated Higgs field. This is specified by using \lstinline|Bar@  H| for the scalar field in \lstinline|AddYukawa|. With the conjugated Higgs field, the $ \SU(2)_\LL $ contraction is done with a Kronecker delta instead of the antisymmetric invariant. 

The quartic Higgs coupling is added to the model with the call
\begin{lstlisting}[firstnumber = 23, mathescape =true]
AddQuartic[$\lambda$, {Bar@ H, H, Bar@ H, H}, 
	GroupInvariant->(del[SU2L@ fund, #1, #2] del[SU2L@ fund, #3, #4] /2 &)]
\end{lstlisting}
Again, the function takes the coupling and the fields (four scalars this time) as arguments. In contrast to the matrix Yukawa couplings of the SM, $ \lambda $ is a scalar coupling, so no coupling indices have to be passed to the function. The $ 1/2 $ normalization of the Higgs self-coupling in Eq.~\eqref{eq:SM_quartic} is put in the group invariant. \rgbeta never assumes any normalization factors even when, as is the case with the SM, there is a symmetry factor associated with the coupling. 

In \msbar the relevant couplings do not influence the running of the marginal couplings. However, if we wish to explore the running of the Higgs mass parameter, we can add it to the model with 
\begin{lstlisting}[firstnumber = 25]
AddScalarMass[M2, {Bar@ H, H}, 
	GroupInvariant->(del[SU2L@ fund, #1, #2] &)]
\end{lstlisting}
completely parallel to the implementation of the other couplings. 
This concludes the implementation of the SM. 

Before proceeding to extract the SM \befs, we would like comment on cases with other invariants in the couplings. In some cases, it is possible to construct new invariants using the group invariants already implemented in \rgbeta. In such cases, the user can manually define the relevant contractions, but it is important to avoid repeating index names in the internal contractions. The best way to avoid this is to use a combination of \lstinline|SetDelayed| and \lstinline|Module| to ensure unique naming of the indices every time the invariant is called. An example is $ \SU(5) $ GUT, where there is an adjoint scalar $ \Sigma $, which has two independent invariants with four fields, one of them being $ (\Sigma_a \Sigma_a)^2  $. For the second, independent invariant, one often uses a trace over 4 fundamental generators, which can be implemented in the program as    
\begin{example}[mathescape, emph={A, B, C, D, A_, B_, C_, D_}]
In[1]	adj4inv[A_, B_, C_, D_ ] := Module[{a, b, c, d}, 
		  tGen[SU5@fund, A, a, b] tGen[SU5@fund, B, b, c] 
		  tGen[SU5@fund, C, c, d] tGen[SU5@fund, D, d, a] ];
In[2]	AddQuartic[$ \lambda\Sigma $, {$ \Sigma $, $ \Sigma $, $ \Sigma $, $ \Sigma $},
		  GroupInvariant -> (adj4inv[#1, #2, #3, #4] &) ]
\end{example}
The full $ \SU(5) $ implementation is included with the package in \path{/Documentation/Sample_Models.nb}.

\subsection{Producing the \befs}
Once the SM model has been loaded into the Mathematica kernel, extracting the \bef is as simple as anything. To obtain, for instance, the 2-loop \bef of the down-type Yukawa coupling, one simply has to call the function \lstinline|BetaFunction[yd, 2]|. This function returns the \bef as defined in Eqs.~(\ref{eq:beta_yuk_quart}--\ref{eq:beta_abelian}). 

A typical use of \lstinline|BetaFunction| (selected for minimality) will look like 
{ \mathversion{subsectionmath}
\begin{example}[mathescape]
In[1]		BetaFunction[ye, 1]
Out[1]		$ \dfrac{1}{16\pi^2} \bigg( \eminus \dfrac{15}{4}$g1$ ^2 $ye$_{\text{\$i,\$j}} $ -$\dfrac{9}{4}$g2$ ^2 $ye$_{\text{\$i,\$j}} $ +3Tr[yd.yd$^{\bm{\dagger}}$]ye$_{\text{\$i,\$j}} $ +Tr[ye.ye$^{\bm{\dagger}}$]ye$_{\text{\$i,\$j}} $ 
				+3Tr[yu.yu$^{\bm{\dagger}}$]ye$_{\text{\$i,\$j}} $ +$ \dfrac{3}{2} $ye.ye$^{\bm{\dagger}}$.ye$_{\text{\$i,\$j}} \bigg)$
\end{example}} 
\noindent for the lepton Yukawa \bef at 1-loop order.
The formatting of the output (in the Mathematica \lstinline|StandardForm|) has been set up to make the output readable to the user. To see the underlying Mathematica expression, one can use the \lstinline|InputForm| command. 
The \lstinline|$i| and \lstinline|$j| indices are the flavor indices carried by the \lstinline|ye| coupling. These correspond to the open indices of the \lstinline|ye| \bef. \lstinline|$i| is used to denote an index of the first fermion of the coupling and \lstinline|$j| is an index of the second fermion (as they are given when specifying the coupling).\footnote{Using \lstinline|InputForm| the user will be able to tell that these indices are both of the \lstinline|gen| type.} With how we implemented the \lstinline|ye| coupling, these are the \lstinline|l| and \lstinline|e| generation indices, respectively. For further manipulation with the \bef, one will typically wish to remove such explicit indices to leave the matrix structure implicit. This can be done with the \lstinline|Finalize| routine, which can also be used to substitute Matrix/vector couplings with a list of their entries. We should mention that \lstinline|BetaTerm| can be used to single out the contribution to a \bef at a particular loop order, $ \beta^{(\ell)}_g $. 

In many models, there will be multiple singlets in the product of the same four scalars allowing for multiple couplings. The quintessential example of this behavior is the single-/double-trace coupling of scalars in the fundamental representation of two different groups. In such events, the naive operators used in \rgbeta to project out specific couplings will mix the couplings in question and \lstinline|BetaTerm|/\lstinline|BetaFunction| cannot be relied on to produce the correct \befs for these couplings. This behavior can also be seen with \lstinline|CheckProjection|, which will mix the couplings in question. In such cases, one should rely on \lstinline|QuarticBetaFunctions|, which extracts all quartic \befs simultaneously and accounts for the mixing by identifying the correct linear combination of the projection operators.   

This concludes our discussion of the functionality of \rgbeta. Examples of further manipulation of the \befs can be found in \path{/Documentation/Tutorial.nb}. Here one can also find an example of how matrix couplings can be parametrized to obtain e.g. the \bef of the charm Yukawa coupling.

\section{Summary and Conclusions}
In recent years there has been a renewed interest in higher-order \befs for model building and precision physics. The theory of the general \bef has also matured to the point where we now have a completely general formalism that treats the couplings in a unified way, correcting several mistakes along the way~\cite{Pickering:2001aq,Schienbein:2018fsw,Poole:2019kcm,Sartore:2020pkk}. \rgbeta is a Mathematica package that leverages these theory advances along with the structure delta approach to Lagrangian matching~\cite{Molgaard:2014hpa} in a minimal yet fairly general tool for the automatic computation of \befs in four-dimensional renormalizable models. 

The \rgbeta package presented in this paper is an attempt at striking a balance between having an intuitive and minimal way of implementing models (the problematic part of using any RG tool) and maintaining a good degree of generality. It is fast enough that it can typically be evaluated directly in the notebook environment, allowing the user to experiment with the resulting \befs. It also has the advantage that group indices can be kept arbitrary. It is our ambition to implement the full set of \befs up to order 4-3-2 in \rgbeta pending ongoing work with J. Davies and F.~Herren~\cite{Herren}.

\subsection*{Acknowledgments}
I would like to thank Lohan Sartore for pointing out the subtleties in the dummy-field method for extracting the \befs for the relevant couplings. Likewise, Florian Herren was very helpful in tracking down the issue in the 2HDM \befs and carefully reading the manuscript. This work has received funding from the Swiss National Science Foundation (SNF) through the Eccellenza Professorial Fellowship ``Flavor Physics at the High Energy Frontier'' project number 186866.

\app[Appendix]
Here we include an overview of the routines of \rgbeta and their options presented in alphabetical order. All options of the routines are given with their default value. 

\begin{function}[emph = {field}]
AddFermion[field, Options]
\end{function}
adds a Weyl fermion field to the model with specified charge and flavor.
\begin{arguments}
	\item \lstinline[emph={field}]|field| is the name of the field.
	\options
	\begin{arguments}
		\item\lstinline|GaugeRep -> {}| is a list of representations under the gauge product groups. 
		\item\lstinline|FlavorIndices -> {}| is a list of the flavor indices and representations of global symmetries of the field. 
		\argend
	\end{arguments} 
\end{arguments} 

\begin{function}[emph={coupling, psi1, psi2}]
AddFermionMass[coupling, {psi1, psi2}, Options]
\end{function}
defines a fermion mass term in the model.
\begin{arguments}
	\item \argument{coupling} names the mass parameter of the mass terms.
	\item \lstinline[emph = {psi1, psi2}]|{psi1, psi2}| is the list consisting of the two fermions \argument{psi1} and \argument{psi2} of the mass term. 
	\options
	\begin{arguments}
		\item \lstinline|GroupInvariant -> (1 &)| is a pure function of 2 arguments defining the group invariants of the coupling. 
		\item \lstinline|MassIndices -> (Null &)| is a pure function of 2 arguments that specifies the tensor indices of the coupling.
		\item \lstinline|Chirality -> Left| sets whether the coupling appears with left-handed or right-handed fields in the Lagrangian.
		\argend
 	\end{arguments}
\end{arguments}

\begin{function}[emph = {coupling, groupName, lieGroup, n}]
AddGaugeGroup[coupling, groupName, lieGroup[n], Options]
\end{function}
adds a gauge group to the current model.
\begin{arguments}
	\item \argument{coupling} specifies the coupling of the gauge group.
	\item \argument{groupName} specifies the reference name associated to the group and its representations.
	\item \lstinline[emph={lieGroup, n}]|lieGroup[n]| specifies what Lie Group is gauged. The options are \lstinline|U1|$ = $\lstinline|U1[1]|, \lstinline|U1[n]|, \lstinline[emph=n]|SU[n]|, \lstinline[emph=n]|Sp[n]|, and \lstinline[emph=n]|SO[n]|, with \argument{n} either an integer or a symbol. While the other group names are self explanatory, \lstinline|U1[n]| is used to denote a $ \U^n(1) $ group.   
	\options
	\begin{arguments}
	\item \lstinline|CouplingMatrix -> Automatic| determines the naming of the coupling matrix if the Lie group  is $ \U^n(1) $. Any symmetric $ n\times n $ matrix can be supplied instead of the automatic naming. 
	\argend
	\end{arguments} 
\end{arguments} 

\begin{function}[emph={coupling, phi1, phi2, phi3, phi4}]
AddQuartic[coupling, {phi1, phi2, phi3, phi4}, Options]
\end{function}
defines a quartic coupling in the model.
\begin{arguments}
	\item \argument{coupling} specifies the coupling constant of the quartic interaction.
	\item \lstinline[emph = {phi1, phi2, phi3, phi4}]|{phi1, phi2, phi3, phi4}| is a list consisting of the four scalar fields involved in the interaction. They can be individually conjugated with \lstinline|Bar|.
	\options
	\begin{arguments}
		\item \lstinline|GroupInvariant -> (1 &)| is a pure function of 4 arguments defining the group invariants of the coupling. 
		\item \lstinline|CouplingIndices -> (Null &)| is a pure function of 4 arguments that specifies the tensor indices of the coupling.
		\item \lstinline|SelfConjugate -> True| sets if the interaction is real, or whether the Hermitian conjugate appears in the Lagrangian as well.
		\argend
	\end{arguments}
\end{arguments}

\begin{function}[emph = {field}]
AddScalar[field, Options]
\end{function}
adds a scalar field to the model with specified charge and flavor.
\begin{arguments}
	\item \lstinline[emph={field}]|field| is the name of the field.
	\options
	\begin{arguments}
		\item\lstinline|GaugeRep -> {}| is a list of representations under the gauge groups. 
		\item\lstinline|FlavorIndices -> {}| is a list of the flavor indices and representations of global symmetries of the field.
		\item\lstinline|SelfConjugate -> False| determines if the fields is complex or real. 
		\argend
	\end{arguments} 
\end{arguments} 

\begin{function}[emph={coupling, phi1, phi2, phi3}]
AddScalarMass[coupling, {phi1, phi2}, Options]
\end{function}
defines a scalar mass term in the model.
\begin{arguments}
	\item \argument{coupling} denotes the mass parameter of the mass term \emph{assumed to have mass-dimension two}.
	\item \lstinline[emph = {phi1, phi2}]|{phi1, phi2}| is the list consisting of two scalar fields. They can be be conjugated individually with \lstinline|Bar|
	\options
	\begin{arguments}
		\item \lstinline|GroupInvariant -> (1 &)| is a pure function of 2 arguments defining the group invariants of the coupling. 
		\item \lstinline|CouplingIndices -> (Null &)| is a pure function of 2 arguments that specifies the tensor indices of the coupling.
		\item \lstinline|SelfConjugate -> True| sets if the interaction is real, or whether the Hermitian conjugate appears in the Lagrangian as well.
		\argend
	\end{arguments}
\end{arguments}

\begin{function}[emph={coupling, phi1, phi2, phi3}]
AddTrilinear[coupling, {phi1, phi2, phi3}, Options]
\end{function}
defines a trilinear scalar coupling in the model.
\begin{arguments}
	\item \argument{coupling} specifies the coupling constant of the trilinear interaction.
	\item \lstinline[emph = {phi1, phi2, phi3}]|{phi1, phi2, phi3}| is the list consisting of three scalar fields. They can be be conjugated with \lstinline|Bar|
	\options
	\begin{arguments}
		\item \lstinline|GroupInvariant -> (1 &)| is a pure function of 3 arguments defining the group invariants of the coupling. 
		\item \lstinline|CouplingIndices -> (Null &)| is a pure function of 3 arguments that specifies the tensor indices of the coupling.
		\item \lstinline|SelfConjugate -> True| sets if the interaction is real, or whether the Hermitian conjugate appears in the Lagrangian as well.
		\argend
	\end{arguments}
\end{arguments}

\begin{function}[emph={coupling, phi, psi1, psi2}]
AddYukawa[coupling, {phi, psi1, psi2}, Options]
\end{function}
defines a Yukawa coupling in the model.
\begin{arguments}
	\item \argument{coupling} specifies the coupling constant of the Yukawa interaction.
	\item \lstinline[emph = {phi, psi1, psi2}]|{phi, psi1, psi2}| is the list consisting of the scalar \argument{phi} and two fermions \argument{psi1} and \argument{psi2} of the interaction. The scalar can be conjugated with \lstinline|Bar|. 
	\options
	\begin{arguments}
	\item \lstinline|GroupInvariant -> (1 &)| is a pure function of 3 arguments defining the group invariants of the coupling. 
	\item \lstinline|CouplingIndices -> (Null &)| is a pure function of 3 arguments that specifies the tensor indices of the coupling.
	\item \lstinline|Chirality -> Left| sets whether the coupling appears with left-handed or right-handed fields in the Lagrangian.
	\argend
 	\end{arguments}
\end{arguments}

\begin{function}[emph = {field, loop}]
AnomalousDimension[field, loop, Options]
\end{function}
gives the full anomalous dimension of a field up to the given loop order.
\begin{arguments}
	\item \argument{field} the field one wishes to find the anomalous dimensions of.
	\item \argument{loop} is either \lstinline|1| or \lstinline|2| specifying to what loop order.
	\options
	\begin{arguments}
		\item \lstinline|RescaledCouplings -> False| determines whether the couplings should all be rescaled with $ g \to 4\pi g $, $ y \to 4\pi y $, and $ \lambda \to (4\pi)^2 \lambda $ in the output.
		\argend
	\end{arguments} 
\end{arguments} 

\begin{function}[emph = {field, loop}]
AnomalousDimTerm[field, loop]
\end{function}
gives the term $ \gamma_\phi^{(\ell)} $ in the anomalous dimension.  
\begin{arguments}
	\item \argument{field} the field one wishes to find the anomalous dimensions of.
	\item \argument{loop} is either \lstinline|1| or \lstinline|2| specifying the loop order.   
	\argend
\end{arguments}

\begin{function}[emph = {coupling, loop}]
BetaFunction[coupling, loop, Options]
\end{function}
gives the full \bef of a coupling up to the given loop order.
\begin{arguments}
	\item \argument{coupling} the coupling one wishes to find the \bef of.
	\item \argument{loop} is an integer specifying to what loop order.
	\options
	\begin{arguments}
		\item \lstinline|FourDimensions -> True| determines if the \bef is strictly four-dimensional or whether it should include the zeroth order $ \ord{\epsilon} $ term in $ (4-\epsilon) $ dimensions.
		\item \lstinline|RescaledCouplings -> False| determines whether the couplings should all be rescaled with $ g \to 4\pi g $, $ y \to 4\pi y $, and $ \lambda \to (4\pi)^2 \lambda $.
		\argend
	\end{arguments} 
\end{arguments} 

\begin{function}[emph = {coupling, loop}]
BetaTerm[coupling, loop]
\end{function}
gives the term $ \beta_g^{(\ell)} $ in the \bef.  
\begin{arguments}
	\item \argument{coupling} is a symbol corresponding to the coupling of the relevant \bef.
	\item \argument{loop} is an integer specifying the loop order.   
	\argend
\end{arguments}

\begin{function}[emph={coupling}]
CheckProjection[coupling]
\end{function}
returns the value of the internal projection operator applied to the general coupling. 
\begin{arguments}
	\item \argument{coupling} specifies the coupling to be checked.
	\argend
\end{arguments}

\begin{function}[emph = {groupName, lieGroup, n}]
DefineLieGroup[groupName, lieGroup[n]]
\end{function}
sets up all group constants associated to the a Lie group of the specified kind. This can be used e.g. if the model contains a global symmetry. 
\begin{arguments}
	\item \argument{groupName} specifies the name of the added gauge group.
	\item \lstinline[emph={lieGroup, n}]|lieGroup[n]| specifies the type Lie Group. It can either be \lstinline[emph=n]|SU[n]|, \lstinline[emph=n]|Sp[n]|, or \lstinline[emph=n]|SO[n]| with \argument{n} either an integer or symbol.
	\argend
\end{arguments} 

\begin{function}[emph = {expr}]
Finalize[expr, Options]
\end{function}
further manipulates the expression (typically a \bef), removing indices unnecessary indices and allowing for couplings substitutions.
\begin{arguments}
	\item \argument{expr} an expression on the form such as given by \lstinline|BetaTerm|.
	\options 
	\begin{arguments}
		\item \lstinline|Parametrizations -> {}| a set of substitution rules, to replace the coupling symbols with e.g. coupling matrices. 
		\item \lstinline|BarToConjugate -> False| replaces the \rgbeta head \lstinline|Bar| with the standard  Mathematica \lstinline|Conjugate|, giving an expression more suitable for further numerical analysis.  
 	\argend
	\end{arguments}
\end{arguments} 

\begin{function}[emph = {loop}]
QuarticBetaFunctions[loop, Options]
\end{function}
gives the full \befs of all quartic couplings using fully diagonalized projectors.
\begin{arguments}
	\item \argument{loop} is an integer specifying to what loop order.
	\options
	\begin{arguments}
		\item \lstinline|FourDimensions -> True| determines if the \bef is strictly four-dimensional or whether it should include the zeroth order $ \ord{\epsilon} $ term in $ (4-\epsilon) $ dimensions.
		\item \lstinline|RescaledCouplings -> False| determines whether the couplings should all be rescaled with $ g \to 4\pi g $, $ y \to 4\pi y $, and $ \lambda \to (4\pi)^2 \lambda $.
		\argend
	\end{arguments} 
\end{arguments} 

\begin{function}
ResetModel[ ]
\end{function}
clears all current model definitions, allowing for defining another model without having to quit the kernel and reloading \rgbeta. \argend

\begin{function}[emph={symbol}]
SetReal[symbol,...]
\end{function}
instructs \rgbeta to treat one or more symbols (typically couplings) as being real by setting \lstinline|Bar@  symb = symb|. 
\begin{arguments}
	\item \argument{symbol} the symbol that will be defined to be real.
	\argend
\end{arguments}

\sectionlike{References}
\vspace{-10pt}
\nocite{apsrev41Control}
\bibliography{References} 	

\begin{thebibliography}{29}%
\makeatletter
\providecommand \@ifxundefined [1]{%
 \@ifx{#1\undefined}
}%
\providecommand \@ifnum [1]{%
 \ifnum #1\expandafter \@firstoftwo
 \else \expandafter \@secondoftwo
 \fi
}%
\providecommand \@ifx [1]{%
 \ifx #1\expandafter \@firstoftwo
 \else \expandafter \@secondoftwo
 \fi
}%
\providecommand \natexlab [1]{#1}%
\providecommand \enquote  [1]{``#1''}%
\providecommand \bibnamefont  [1]{#1}%
\providecommand \bibfnamefont [1]{#1}%
\providecommand \citenamefont [1]{#1}%
\providecommand \href@noop [0]{\@secondoftwo}%
\providecommand \href [0]{\begingroup \@sanitize@url \@href}%
\providecommand \@href[1]{\@@startlink{#1}\@@href}%
\providecommand \@@href[1]{\endgroup#1\@@endlink}%
\providecommand \@sanitize@url [0]{\catcode `\\12\catcode `\$12\catcode
  `\&12\catcode `\#12\catcode `\^12\catcode `\_12\catcode `\%12\relax}%
\providecommand \@@startlink[1]{}%
\providecommand \@@endlink[0]{}%
\providecommand \url  [0]{\begingroup\@sanitize@url \@url }%
\providecommand \@url [1]{\endgroup\@href {#1}{\urlprefix }}%
\providecommand \urlprefix  [0]{URL }%
\providecommand \Eprint [0]{\href }%
\providecommand \doibase [0]{http://dx.doi.org/}%
\providecommand \selectlanguage [0]{\@gobble}%
\providecommand \bibinfo  [0]{\@secondoftwo}%
\providecommand \bibfield  [0]{\@secondoftwo}%
\providecommand \translation [1]{[#1]}%
\providecommand \BibitemOpen [0]{}%
\providecommand \bibitemStop [0]{}%
\providecommand \bibitemNoStop [0]{.\EOS\space}%
\providecommand \EOS [0]{\spacefactor3000\relax}%
\providecommand \BibitemShut  [1]{\csname bibitem#1\endcsname}%
\let\auto@bib@innerbib\@empty
\bibitem [{\citenamefont {Giudice}\ \emph {et~al.}(2015)\citenamefont
  {Giudice}, \citenamefont {Isidori}, \citenamefont {Salvio},\ and\
  \citenamefont {Strumia}}]{Giudice:2014tma}%
  \BibitemOpen
  \bibfield  {author} {\bibinfo {author} {\bibfnamefont {G.~F.}\ \bibnamefont
  {Giudice}}, \bibinfo {author} {\bibfnamefont {G.}~\bibnamefont {Isidori}},
  \bibinfo {author} {\bibfnamefont {A.}~\bibnamefont {Salvio}}, \ and\ \bibinfo
  {author} {\bibfnamefont {A.}~\bibnamefont {Strumia}},\ }\bibfield  {title}
  {\enquote {\bibinfo {title} {{Softened Gravity and the Extension of the
  Standard Model up to Infinite Energy}},}\ }\href {\doibase
  10.1007/JHEP02(2015)137} {\bibfield  {journal} {\bibinfo  {journal} {JHEP}\
  }\textbf {\bibinfo {volume} {02}},\ \bibinfo {pages} {137} (\bibinfo {year}
  {2015})},\ \Eprint {http://arxiv.org/abs/1412.2769} {arXiv:1412.2769
  [hep-ph]}\BibitemShut {NoStop}%
\bibitem [{\citenamefont {Litim}\ and\ \citenamefont
  {Sannino}(2014)}]{Litim:2014uca}%
  \BibitemOpen
  \bibfield  {author} {\bibinfo {author} {\bibfnamefont {D.~F.}\ \bibnamefont
  {Litim}}\ and\ \bibinfo {author} {\bibfnamefont {F.}~\bibnamefont
  {Sannino}},\ }\bibfield  {title} {\enquote {\bibinfo {title} {{Asymptotic
  safety guaranteed}},}\ }\href {\doibase 10.1007/JHEP12(2014)178} {\bibfield
  {journal} {\bibinfo  {journal} {JHEP}\ }\textbf {\bibinfo {volume} {12}},\
  \bibinfo {pages} {178} (\bibinfo {year} {2014})},\ \Eprint
  {http://arxiv.org/abs/1406.2337} {arXiv:1406.2337 [hep-th]}\BibitemShut
  {NoStop}%
\bibitem [{\citenamefont {Machacek}\ and\ \citenamefont
  {Vaughn}(1983)}]{Machacek:1983tz}%
  \BibitemOpen
  \bibfield  {author} {\bibinfo {author} {\bibfnamefont {M.~E.}\ \bibnamefont
  {Machacek}}\ and\ \bibinfo {author} {\bibfnamefont {M.~T.}\ \bibnamefont
  {Vaughn}},\ }\bibfield  {title} {\enquote {\bibinfo {title} {{Two Loop
  Renormalization Group Equations in a General Quantum Field Theory. 1. Wave
  Function Renormalization}},}\ }\href {\doibase 10.1016/0550-3213(83)90610-7}
  {\bibfield  {journal} {\bibinfo  {journal} {Nucl. Phys.}\ }\textbf {\bibinfo
  {volume} {B222}},\ \bibinfo {pages} {83--103} (\bibinfo {year}
  {1983})}\BibitemShut {NoStop}%
\bibitem [{\citenamefont {Machacek}\ and\ \citenamefont
  {Vaughn}(1984)}]{Machacek:1983fi}%
  \BibitemOpen
  \bibfield  {author} {\bibinfo {author} {\bibfnamefont {M.~E.}\ \bibnamefont
  {Machacek}}\ and\ \bibinfo {author} {\bibfnamefont {M.~T.}\ \bibnamefont
  {Vaughn}},\ }\bibfield  {title} {\enquote {\bibinfo {title} {{Two Loop
  Renormalization Group Equations in a General Quantum Field Theory. 2. Yukawa
  Couplings}},}\ }\href {\doibase 10.1016/0550-3213(84)90533-9} {\bibfield
  {journal} {\bibinfo  {journal} {Nucl. Phys.}\ }\textbf {\bibinfo {volume}
  {B236}},\ \bibinfo {pages} {221--232} (\bibinfo {year} {1984})}\BibitemShut
  {NoStop}%
\bibitem [{\citenamefont {Machacek}\ and\ \citenamefont
  {Vaughn}(1985)}]{Machacek:1984zw}%
  \BibitemOpen
  \bibfield  {author} {\bibinfo {author} {\bibfnamefont {M.~E.}\ \bibnamefont
  {Machacek}}\ and\ \bibinfo {author} {\bibfnamefont {M.~T.}\ \bibnamefont
  {Vaughn}},\ }\bibfield  {title} {\enquote {\bibinfo {title} {{Two Loop
  Renormalization Group Equations in a General Quantum Field Theory. 3. Scalar
  Quartic Couplings}},}\ }\href {\doibase 10.1016/0550-3213(85)90040-9}
  {\bibfield  {journal} {\bibinfo  {journal} {Nucl. Phys.}\ }\textbf {\bibinfo
  {volume} {B249}},\ \bibinfo {pages} {70--92} (\bibinfo {year}
  {1985})}\BibitemShut {NoStop}%
\bibitem [{\citenamefont {Jack}\ and\ \citenamefont
  {Osborn}(1985)}]{Jack:1984vj}%
  \BibitemOpen
  \bibfield  {author} {\bibinfo {author} {\bibfnamefont {I.}~\bibnamefont
  {Jack}}\ and\ \bibinfo {author} {\bibfnamefont {H.}~\bibnamefont {Osborn}},\
  }\bibfield  {title} {\enquote {\bibinfo {title} {{General Background Field
  Calculations With Fermion Fields}},}\ }\href {\doibase
  10.1016/0550-3213(85)90088-4} {\bibfield  {journal} {\bibinfo  {journal}
  {Nucl. Phys. B}\ }\textbf {\bibinfo {volume} {249}},\ \bibinfo {pages}
  {472--506} (\bibinfo {year} {1985})}\BibitemShut {NoStop}%
\bibitem [{\citenamefont {Pickering}\ \emph {et~al.}(2001)\citenamefont
  {Pickering}, \citenamefont {Gracey},\ and\ \citenamefont
  {Jones}}]{Pickering:2001aq}%
  \BibitemOpen
  \bibfield  {author} {\bibinfo {author} {\bibfnamefont {A.~G.~M.}\
  \bibnamefont {Pickering}}, \bibinfo {author} {\bibfnamefont {J.~A.}\
  \bibnamefont {Gracey}}, \ and\ \bibinfo {author} {\bibfnamefont {D.~R.~T.}\
  \bibnamefont {Jones}},\ }\bibfield  {title} {\enquote {\bibinfo {title}
  {{Three loop gauge beta function for the most general single gauge coupling
  theory}},}\ }\href {\doibase 10.1016/S0370-2693(02)01779-3,
  10.1016/S0370-2693(01)00624-4} {\bibfield  {journal} {\bibinfo  {journal}
  {Phys. Lett.}\ }\textbf {\bibinfo {volume} {B510}},\ \bibinfo {pages}
  {347--354} (\bibinfo {year} {2001})},\ \bibinfo {note} {[Erratum: Phys.
  Lett.B535,377(2002)]},\ \Eprint {http://arxiv.org/abs/hep-ph/0104247}
  {arXiv:hep-ph/0104247 [hep-ph]}\BibitemShut {NoStop}%
\bibitem [{\citenamefont {Staub}(2008)}]{Staub:2008uz}%
  \BibitemOpen
  \bibfield  {author} {\bibinfo {author} {\bibfnamefont {F.}~\bibnamefont
  {Staub}},\ }\bibfield  {title} {\enquote {\bibinfo {title} {{SARAH}},}\
  }\href@noop {} {\  (\bibinfo {year} {2008})},\ \Eprint
  {http://arxiv.org/abs/0806.0538} {arXiv:0806.0538 [hep-ph]}\BibitemShut
  {NoStop}%
\bibitem [{\citenamefont {Staub}(2014)}]{Staub:2013tta}%
  \BibitemOpen
  \bibfield  {author} {\bibinfo {author} {\bibfnamefont {F.}~\bibnamefont
  {Staub}},\ }\bibfield  {title} {\enquote {\bibinfo {title} {{SARAH 4: A tool
  for (not only SUSY) model builders}},}\ }\href {\doibase
  10.1016/j.cpc.2014.02.018} {\bibfield  {journal} {\bibinfo  {journal}
  {Comput. Phys. Commun.}\ }\textbf {\bibinfo {volume} {185}},\ \bibinfo
  {pages} {1773--1790} (\bibinfo {year} {2014})},\ \Eprint
  {http://arxiv.org/abs/1309.7223} {arXiv:1309.7223 [hep-ph]}\BibitemShut
  {NoStop}%
\bibitem [{\citenamefont {Luo}\ \emph {et~al.}(2003)\citenamefont {Luo},
  \citenamefont {Wang},\ and\ \citenamefont {Xiao}}]{Luo:2002ti}%
  \BibitemOpen
  \bibfield  {author} {\bibinfo {author} {\bibfnamefont {M.-x.}\ \bibnamefont
  {Luo}}, \bibinfo {author} {\bibfnamefont {H.-w.}\ \bibnamefont {Wang}}, \
  and\ \bibinfo {author} {\bibfnamefont {Y.}~\bibnamefont {Xiao}},\ }\bibfield
  {title} {\enquote {\bibinfo {title} {{Two loop renormalization group
  equations in general gauge field theories}},}\ }\href {\doibase
  10.1103/PhysRevD.67.065019} {\bibfield  {journal} {\bibinfo  {journal} {Phys.
  Rev.}\ }\textbf {\bibinfo {volume} {D67}},\ \bibinfo {pages} {065019}
  (\bibinfo {year} {2003})},\ \Eprint {http://arxiv.org/abs/hep-ph/0211440}
  {arXiv:hep-ph/0211440 [hep-ph]}\BibitemShut {NoStop}%
\bibitem [{\citenamefont {Sartore}\ and\ \citenamefont
  {Schienbein}(2020)}]{Sartore:2020gou}%
  \BibitemOpen
  \bibfield  {author} {\bibinfo {author} {\bibfnamefont {L.}~\bibnamefont
  {Sartore}}\ and\ \bibinfo {author} {\bibfnamefont {I.}~\bibnamefont
  {Schienbein}},\ }\bibfield  {title} {\enquote {\bibinfo {title} {{PyR@TE
  3}},}\ }\href@noop {} {\  (\bibinfo {year} {2020})},\ \Eprint
  {http://arxiv.org/abs/2007.12700} {arXiv:2007.12700 [hep-ph]}\BibitemShut
  {NoStop}%
\bibitem [{\citenamefont {Lyonnet}\ \emph {et~al.}(2014)\citenamefont
  {Lyonnet}, \citenamefont {Schienbein}, \citenamefont {Staub},\ and\
  \citenamefont {Wingerter}}]{Lyonnet:2013dna}%
  \BibitemOpen
  \bibfield  {author} {\bibinfo {author} {\bibfnamefont {F.}~\bibnamefont
  {Lyonnet}}, \bibinfo {author} {\bibfnamefont {I.}~\bibnamefont {Schienbein}},
  \bibinfo {author} {\bibfnamefont {F.}~\bibnamefont {Staub}}, \ and\ \bibinfo
  {author} {\bibfnamefont {A.}~\bibnamefont {Wingerter}},\ }\bibfield  {title}
  {\enquote {\bibinfo {title} {{PyR@TE: Renormalization Group Equations for
  General Gauge Theories}},}\ }\href {\doibase 10.1016/j.cpc.2013.12.002}
  {\bibfield  {journal} {\bibinfo  {journal} {Comput. Phys. Commun.}\ }\textbf
  {\bibinfo {volume} {185}},\ \bibinfo {pages} {1130--1152} (\bibinfo {year}
  {2014})},\ \Eprint {http://arxiv.org/abs/1309.7030} {arXiv:1309.7030
  [hep-ph]}\BibitemShut {NoStop}%
\bibitem [{\citenamefont {Lyonnet}\ and\ \citenamefont
  {Schienbein}(2017)}]{Lyonnet:2016xiz}%
  \BibitemOpen
  \bibfield  {author} {\bibinfo {author} {\bibfnamefont {F.}~\bibnamefont
  {Lyonnet}}\ and\ \bibinfo {author} {\bibfnamefont {I.}~\bibnamefont
  {Schienbein}},\ }\bibfield  {title} {\enquote {\bibinfo {title} {{PyR@TE 2: A
  Python tool for computing RGEs at two-loop}},}\ }\href {\doibase
  10.1016/j.cpc.2016.12.003} {\bibfield  {journal} {\bibinfo  {journal}
  {Comput. Phys. Commun.}\ }\textbf {\bibinfo {volume} {213}},\ \bibinfo
  {pages} {181--196} (\bibinfo {year} {2017})},\ \Eprint
  {http://arxiv.org/abs/1608.07274} {arXiv:1608.07274 [hep-ph]}\BibitemShut
  {NoStop}%
\bibitem [{\citenamefont {Poole}\ and\ \citenamefont
  {Thomsen}(2019)}]{Poole:2019kcm}%
  \BibitemOpen
  \bibfield  {author} {\bibinfo {author} {\bibfnamefont {C.}~\bibnamefont
  {Poole}}\ and\ \bibinfo {author} {\bibfnamefont {A.~E.}\ \bibnamefont
  {Thomsen}},\ }\bibfield  {title} {\enquote {\bibinfo {title} {{Constraints on
  3- and 4-loop $\beta$-functions in a general four-dimensional Quantum Field
  Theory}},}\ }\href {\doibase 10.1007/JHEP09(2019)055} {\bibfield  {journal}
  {\bibinfo  {journal} {JHEP}\ }\textbf {\bibinfo {volume} {09}},\ \bibinfo
  {pages} {055} (\bibinfo {year} {2019})},\ \Eprint
  {http://arxiv.org/abs/1906.04625} {arXiv:1906.04625 [hep-th]}\BibitemShut
  {NoStop}%
\bibitem [{\citenamefont {Sartore}(2020)}]{Sartore:2020pkk}%
  \BibitemOpen
  \bibfield  {author} {\bibinfo {author} {\bibfnamefont {L.}~\bibnamefont
  {Sartore}},\ }\bibfield  {title} {\enquote {\bibinfo {title} {{General RGEs
  for dimensionful couplings in the $\overline{\mathrm{MS}}$ scheme}},}\ }\href
  {\doibase 10.1103/PhysRevD.102.076002} {\bibfield  {journal} {\bibinfo
  {journal} {Phys. Rev. D}\ }\textbf {\bibinfo {volume} {102}},\ \bibinfo
  {pages} {076002} (\bibinfo {year} {2020})},\ \Eprint
  {http://arxiv.org/abs/2006.12307} {arXiv:2006.12307 [hep-ph]}\BibitemShut
  {NoStop}%
\bibitem [{\citenamefont {Litim}\ and\ \citenamefont
  {Steudtner}(2020)}]{Litim:2020jvl}%
  \BibitemOpen
  \bibfield  {author} {\bibinfo {author} {\bibfnamefont {D.~F.}\ \bibnamefont
  {Litim}}\ and\ \bibinfo {author} {\bibfnamefont {T.}~\bibnamefont
  {Steudtner}},\ }\bibfield  {title} {\enquote {\bibinfo {title} {{ARGES --
  Advanced Renormalisation Group Equation Simplifier}},}\ }\href@noop {} {\
  (\bibinfo {year} {2020})},\ \Eprint {http://arxiv.org/abs/2012.12955}
  {arXiv:2012.12955 [hep-ph]}\BibitemShut {NoStop}%
\bibitem [{\citenamefont {Deppisch}\ and\ \citenamefont
  {Herren}(2020)}]{Deppisch:2020aoj}%
  \BibitemOpen
  \bibfield  {author} {\bibinfo {author} {\bibfnamefont {T.}~\bibnamefont
  {Deppisch}}\ and\ \bibinfo {author} {\bibfnamefont {F.}~\bibnamefont
  {Herren}},\ }\bibfield  {title} {\enquote {\bibinfo {title}
  {{$\texttt{RGE++}:$ A $\texttt{C++}$ library to solve renormalisation group
  equations in quantum field theory}},}\ }\href@noop {} {\  (\bibinfo {year}
  {2020})},\ \Eprint {http://arxiv.org/abs/2101.00021} {arXiv:2101.00021
  [hep-ph]}\BibitemShut {NoStop}%
\bibitem [{\citenamefont {Davies}\ \emph {et~al.}(Work in
  progress)\citenamefont {Davies}, \citenamefont {Herren},\ and\ \citenamefont
  {Thomsen}}]{Herren}%
  \BibitemOpen
  \bibfield  {author} {\bibinfo {author} {\bibfnamefont {J.}~\bibnamefont
  {Davies}}, \bibinfo {author} {\bibfnamefont {F.}~\bibnamefont {Herren}}, \
  and\ \bibinfo {author} {\bibfnamefont {A.~E.}\ \bibnamefont {Thomsen}},\
  }\href@noop {} {} (\bibinfo {year} {Work in progress})\BibitemShut {NoStop}%
\bibitem [{\citenamefont {Jack}\ and\ \citenamefont
  {Osborn}(2014)}]{Jack:2013sha}%
  \BibitemOpen
  \bibfield  {author} {\bibinfo {author} {\bibfnamefont {I.}~\bibnamefont
  {Jack}}\ and\ \bibinfo {author} {\bibfnamefont {H.}~\bibnamefont {Osborn}},\
  }\bibfield  {title} {\enquote {\bibinfo {title} {{Constraints on RG Flow for
  Four Dimensional Quantum Field Theories}},}\ }\href {\doibase
  10.1016/j.nuclphysb.2014.03.018} {\bibfield  {journal} {\bibinfo  {journal}
  {Nucl. Phys. B}\ }\textbf {\bibinfo {volume} {883}},\ \bibinfo {pages}
  {425--500} (\bibinfo {year} {2014})},\ \Eprint
  {http://arxiv.org/abs/1312.0428} {arXiv:1312.0428 [hep-th]}\BibitemShut
  {NoStop}%
\bibitem [{\citenamefont {Antipin}\ \emph {et~al.}(2013)\citenamefont
  {Antipin}, \citenamefont {Gillioz}, \citenamefont {Krog}, \citenamefont
  {M\o{}lgaard},\ and\ \citenamefont {Sannino}}]{Antipin:2013sga}%
  \BibitemOpen
  \bibfield  {author} {\bibinfo {author} {\bibfnamefont {O.}~\bibnamefont
  {Antipin}}, \bibinfo {author} {\bibfnamefont {M.}~\bibnamefont {Gillioz}},
  \bibinfo {author} {\bibfnamefont {J.}~\bibnamefont {Krog}}, \bibinfo {author}
  {\bibfnamefont {E.}~\bibnamefont {M\o{}lgaard}}, \ and\ \bibinfo {author}
  {\bibfnamefont {F.}~\bibnamefont {Sannino}},\ }\bibfield  {title} {\enquote
  {\bibinfo {title} {{Standard Model Vacuum Stability and Weyl Consistency
  Conditions}},}\ }\href {\doibase 10.1007/JHEP08(2013)034} {\bibfield
  {journal} {\bibinfo  {journal} {JHEP}\ }\textbf {\bibinfo {volume} {08}},\
  \bibinfo {pages} {034} (\bibinfo {year} {2013})},\ \Eprint
  {http://arxiv.org/abs/1306.3234} {arXiv:1306.3234 [hep-ph]}\BibitemShut
  {NoStop}%
\bibitem [{\citenamefont {Jack}\ and\ \citenamefont
  {Poole}(2015)}]{Jack:2014pua}%
  \BibitemOpen
  \bibfield  {author} {\bibinfo {author} {\bibfnamefont {I.}~\bibnamefont
  {Jack}}\ and\ \bibinfo {author} {\bibfnamefont {C.}~\bibnamefont {Poole}},\
  }\bibfield  {title} {\enquote {\bibinfo {title} {{The a-function for gauge
  theories}},}\ }\href {\doibase 10.1007/JHEP01(2015)138} {\bibfield  {journal}
  {\bibinfo  {journal} {JHEP}\ }\textbf {\bibinfo {volume} {01}},\ \bibinfo
  {pages} {138} (\bibinfo {year} {2015})},\ \Eprint
  {http://arxiv.org/abs/1411.1301} {arXiv:1411.1301 [hep-th]}\BibitemShut
  {NoStop}%
\bibitem [{\citenamefont {Mihaila}\ \emph {et~al.}(2012)\citenamefont
  {Mihaila}, \citenamefont {Salomon},\ and\ \citenamefont
  {Steinhauser}}]{Mihaila:2012pz}%
  \BibitemOpen
  \bibfield  {author} {\bibinfo {author} {\bibfnamefont {L.~N.}\ \bibnamefont
  {Mihaila}}, \bibinfo {author} {\bibfnamefont {J.}~\bibnamefont {Salomon}}, \
  and\ \bibinfo {author} {\bibfnamefont {M.}~\bibnamefont {Steinhauser}},\
  }\bibfield  {title} {\enquote {\bibinfo {title} {{Renormalization constants
  and beta functions for the gauge couplings of the Standard Model to
  three-loop order}},}\ }\href {\doibase 10.1103/PhysRevD.86.096008} {\bibfield
   {journal} {\bibinfo  {journal} {Phys. Rev. D}\ }\textbf {\bibinfo {volume}
  {86}},\ \bibinfo {pages} {096008} (\bibinfo {year} {2012})},\ \Eprint
  {http://arxiv.org/abs/1208.3357} {arXiv:1208.3357 [hep-ph]}\BibitemShut
  {NoStop}%
\bibitem [{\citenamefont {Mølgaard}(2014)}]{Molgaard:2014hpa}%
  \BibitemOpen
  \bibfield  {author} {\bibinfo {author} {\bibfnamefont {E.}~\bibnamefont
  {Mølgaard}},\ }\bibfield  {title} {\enquote {\bibinfo {title} {{Decrypting
  gauge-Yukawa cookbooks}},}\ }\href {\doibase 10.1140/epjp/i2014-14159-2}
  {\bibfield  {journal} {\bibinfo  {journal} {Eur. Phys. J. Plus}\ }\textbf
  {\bibinfo {volume} {129}},\ \bibinfo {pages} {159} (\bibinfo {year}
  {2014})},\ \Eprint {http://arxiv.org/abs/1404.5550} {arXiv:1404.5550
  [hep-th]}\BibitemShut {NoStop}%
\bibitem [{\citenamefont {Luo}\ and\ \citenamefont {Xiao}(2003)}]{Luo:2002iq}%
  \BibitemOpen
  \bibfield  {author} {\bibinfo {author} {\bibfnamefont {M.-x.}\ \bibnamefont
  {Luo}}\ and\ \bibinfo {author} {\bibfnamefont {Y.}~\bibnamefont {Xiao}},\
  }\bibfield  {title} {\enquote {\bibinfo {title} {{Renormalization group
  equations in gauge theories with multiple U(1) groups}},}\ }\href {\doibase
  10.1016/S0370-2693(03)00076-5} {\bibfield  {journal} {\bibinfo  {journal}
  {Phys. Lett. B}\ }\textbf {\bibinfo {volume} {555}},\ \bibinfo {pages}
  {279--286} (\bibinfo {year} {2003})},\ \Eprint
  {http://arxiv.org/abs/hep-ph/0212152} {arXiv:hep-ph/0212152}\BibitemShut
  {NoStop}%
\bibitem [{\citenamefont {Fonseca}\ \emph {et~al.}(2013)\citenamefont
  {Fonseca}, \citenamefont {Malinsk\'y},\ and\ \citenamefont
  {Staub}}]{Fonseca:2013bua}%
  \BibitemOpen
  \bibfield  {author} {\bibinfo {author} {\bibfnamefont {R.~M.}\ \bibnamefont
  {Fonseca}}, \bibinfo {author} {\bibfnamefont {M.}~\bibnamefont {Malinsk\'y}},
  \ and\ \bibinfo {author} {\bibfnamefont {F.}~\bibnamefont {Staub}},\
  }\bibfield  {title} {\enquote {\bibinfo {title} {{Renormalization group
  equations and matching in a general quantum field theory with kinetic
  mixing}},}\ }\href {\doibase 10.1016/j.physletb.2013.09.042} {\bibfield
  {journal} {\bibinfo  {journal} {Phys. Lett. B}\ }\textbf {\bibinfo {volume}
  {726}},\ \bibinfo {pages} {882--886} (\bibinfo {year} {2013})},\ \Eprint
  {http://arxiv.org/abs/1308.1674} {arXiv:1308.1674 [hep-ph]}\BibitemShut
  {NoStop}%
\bibitem [{\citenamefont {Schienbein}\ \emph {et~al.}(2019)\citenamefont
  {Schienbein}, \citenamefont {Staub}, \citenamefont {Steudtner},\ and\
  \citenamefont {Svirina}}]{Schienbein:2018fsw}%
  \BibitemOpen
  \bibfield  {author} {\bibinfo {author} {\bibfnamefont {I.}~\bibnamefont
  {Schienbein}}, \bibinfo {author} {\bibfnamefont {F.}~\bibnamefont {Staub}},
  \bibinfo {author} {\bibfnamefont {T.}~\bibnamefont {Steudtner}}, \ and\
  \bibinfo {author} {\bibfnamefont {K.}~\bibnamefont {Svirina}},\ }\bibfield
  {title} {\enquote {\bibinfo {title} {{Revisiting RGEs for general gauge
  theories}},}\ }\href {\doibase 10.1016/j.nuclphysb.2018.12.001} {\bibfield
  {journal} {\bibinfo  {journal} {Nucl. Phys. B}\ }\textbf {\bibinfo {volume}
  {939}},\ \bibinfo {pages} {1--48} (\bibinfo {year} {2019})},\ \Eprint
  {http://arxiv.org/abs/1809.06797} {arXiv:1809.06797 [hep-ph]}\BibitemShut
  {NoStop}%
\bibitem [{\citenamefont {Herren}\ \emph {et~al.}(2018)\citenamefont {Herren},
  \citenamefont {Mihaila},\ and\ \citenamefont {Steinhauser}}]{Herren:2017uxn}%
  \BibitemOpen
  \bibfield  {author} {\bibinfo {author} {\bibfnamefont {F.}~\bibnamefont
  {Herren}}, \bibinfo {author} {\bibfnamefont {L.}~\bibnamefont {Mihaila}}, \
  and\ \bibinfo {author} {\bibfnamefont {M.}~\bibnamefont {Steinhauser}},\
  }\bibfield  {title} {\enquote {\bibinfo {title} {{Gauge and Yukawa coupling
  beta functions of two-Higgs-doublet models to three-loop order}},}\ }\href
  {\doibase 10.1103/PhysRevD.97.015016} {\bibfield  {journal} {\bibinfo
  {journal} {Phys. Rev.}\ }\textbf {\bibinfo {volume} {D97}},\ \bibinfo {pages}
  {015016} (\bibinfo {year} {2018})},\ \Eprint
  {http://arxiv.org/abs/1712.06614} {arXiv:1712.06614 [hep-ph]}\BibitemShut
  {NoStop}%
\bibitem [{\citenamefont {Bednyakov}\ \emph {et~al.}(2014)\citenamefont
  {Bednyakov}, \citenamefont {Pikelner},\ and\ \citenamefont
  {Velizhanin}}]{Bednyakov:2013cpa}%
  \BibitemOpen
  \bibfield  {author} {\bibinfo {author} {\bibfnamefont {A.~V.}\ \bibnamefont
  {Bednyakov}}, \bibinfo {author} {\bibfnamefont {A.~F.}\ \bibnamefont
  {Pikelner}}, \ and\ \bibinfo {author} {\bibfnamefont {V.~N.}\ \bibnamefont
  {Velizhanin}},\ }\bibfield  {title} {\enquote {\bibinfo {title} {{Three-loop
  Higgs self-coupling beta-function in the Standard Model with complex Yukawa
  matrices}},}\ }\href {\doibase 10.1016/j.nuclphysb.2013.12.012} {\bibfield
  {journal} {\bibinfo  {journal} {Nucl. Phys.}\ }\textbf {\bibinfo {volume}
  {B879}},\ \bibinfo {pages} {256--267} (\bibinfo {year} {2014})},\ \Eprint
  {http://arxiv.org/abs/1310.3806} {arXiv:1310.3806 [hep-ph]}\BibitemShut
  {NoStop}%
\bibitem [{\citenamefont {Basso}\ \emph {et~al.}(2010)\citenamefont {Basso},
  \citenamefont {Moretti},\ and\ \citenamefont {Pruna}}]{Basso:2010jm}%
  \BibitemOpen
  \bibfield  {author} {\bibinfo {author} {\bibfnamefont {L.}~\bibnamefont
  {Basso}}, \bibinfo {author} {\bibfnamefont {S.}~\bibnamefont {Moretti}}, \
  and\ \bibinfo {author} {\bibfnamefont {G.~M.}\ \bibnamefont {Pruna}},\
  }\bibfield  {title} {\enquote {\bibinfo {title} {{A Renormalisation Group
  Equation Study of the Scalar Sector of the Minimal B-L Extension of the
  Standard Model}},}\ }\href {\doibase 10.1103/PhysRevD.82.055018} {\bibfield
  {journal} {\bibinfo  {journal} {Phys. Rev. D}\ }\textbf {\bibinfo {volume}
  {82}},\ \bibinfo {pages} {055018} (\bibinfo {year} {2010})},\ \Eprint
  {http://arxiv.org/abs/1004.3039} {arXiv:1004.3039 [hep-ph]}\BibitemShut
  {NoStop}%
\end{thebibliography}%
\end{document}